\documentclass[twocolumn,aps,prb,showpacs]{revtex4}

\usepackage{graphicx}
\usepackage{amsmath}
\usepackage{bm}
\usepackage{latexsym,amssymb,float}
\usepackage{setspace}
\allowdisplaybreaks[1]

\newcommand{\nc}{\newcommand}
\nc{\rnc}{\renewcommand}
\nc{\bdm}{\begin{displaymath}}
\nc{\edm}{\end{displaymath}}
\nc{\beq}{\begin{equation}}
\nc{\eeq}{\end{equation}}
\nc{\bea}{\begin{eqnarray}}
\nc{\eea}{\end{eqnarray}}
\nc{\bse}{\begin{subequations}}
\nc{\ese}{\end{subequations}}
\nc{\bpi}{\begin{picture}}
\nc{\epi}{\end{picture}}
\nc{\ba}{\begin{array}}
\nc{\ea}{\end{array}}
\nc{\bsm}{\begin{smallmatrix}}
\nc{\esm}{\end{smallmatrix}}
\nc{\bpm}{\begin{pmatrix}}
\nc{\epm}{\end{pmatrix}}
\nc{\nn}{\nonumber}
\nc{\ds}{\displaystyle}
\nc{\ts}{\textstyle}
\nc{\scs}{\scriptstyle}
\nc{\ltap}{\;\raisebox{-.4ex}{\rlap{$\sim$}}\raisebox{.4ex}{$<$}\;}
\nc{\gtap}{\;\raisebox{-.4ex}{\rlap{$\sim$}}\raisebox{.4ex}{$>$}\;}

\nc{\p}{\partial}
\nc{\ra}{\rightarrow}
\nc{\ua}{\uparrow}
\nc{\da}{\downarrow}
\nc{\dg}{\dagger}
\nc{\f}[2]{\frac{#1}{#2}}

\nc{\al}{\alpha}
\nc{\be}{\beta}
\nc{\ga}{\gamma}
\nc{\de}{\delta}
\nc{\ep}{\epsilon}
\nc{\ka}{\kappa}
\nc{\si}{\sigma}
\nc{\De}{\Delta}

\nc{\Ml}{M_\parallel}
\nc{\Dz}{\De_0}
\nc{\Z}{Z}

\nc{\en}{\phantom{n}}

\nc{\ff}{f}
\nc{\uu}{u}
\nc{\vv}{v}
\nc{\ww}{w}
\nc{\xx}{x}
\nc{\fast}{f^\ast}
\nc{\uast}{u^\ast}
\nc{\vast}{v^\ast}
\nc{\wast}{w^\ast}
\nc{\xast}{x^\ast}

\nc{\eprint}[2]{#1/#2}
\nc{\ibid}[3]{{\em ibid.} {\bf #1}, #2 (#3)}
\nc{\APNY}[3]{Ann.\ Phys.\ (N.Y.) \textbf{#1}, #2 (#3)}
\nc{\CPC}[3]{Comp.\ Phys.\ Comm.\ \textbf{#1}, #2 (#3)}
\nc{\EPJB}[3]{Eur.\ Phys.\ J.\ B \textbf{#1}, #2 (#3)}
\nc{\EPL}[3]{Europhys.\ Lett.\ \textbf{#1}, #2 (#3)}
\nc{\IJMPB}[3]{Int.\ J.\ Mod.\ Phys.\ B \textbf{#1}, #2 (#3)}
\nc{\JChP}[3]{J.\ Chem.\ Phys.\ \textbf{#1}, #2 (#3)}
\nc{\JCpP}[3]{J.\ Comput.\ Phys.\ \textbf{#1}, #2 (#3)}
\nc{\JHEP}[3]{JHEP \textbf{#1}, #2 (#3)}
\nc{\JMP}[3]{J.\ Math.\ Phys.\ \textbf{#1}, #2 (#3)}
\nc{\JPB}[3]{J.\ Phys.\ B \textbf{#1}, #2 (#3)}
\nc{\JPF}[3]{J.\ Phys.\ (France) \textbf{#1}, #2 (#3)}
\nc{\JTB}[3]{J.\ Theor.\ Biol.\ \textbf{#1}, #2 (#3)}
\nc{\LP}[3]{Laser Phys.\ \textbf{#1}, #2 (#3)}
\nc{\MPLB}[3]{Mod.\ Phys.\ Lett.\ B \textbf{#1}, #2 (#3)}
\nc{\MUPB}[3]{Moscow Univ.\ Phys.\ Bull.\ \textbf{#1}, #2 (#3)}
\nc{\NCD}[3]{Nuovo Cimento D \textbf{#1}, #2 (#3)}
\nc{\NCL}[3]{Nuovo Cimento Lett.\ \textbf{#1}, #2 (#3)}
\nc{\NPB}[3]{Nucl.\ Phys. B \textbf{#1}, #2 (#3)}
\nc{\PA}[3]{Physica A \textbf{#1}, #2 (#3)}
\nc{\PLA}[3]{Phys.\ Lett.\ A \textbf{#1}, #2 (#3)}
\nc{\PLB}[3]{Phys.\ Lett.\ B \textbf{#1}, #2 (#3)}
\nc{\PR}[3]{Phys.\ Rev.\ \textbf{#1}, #2 (#3)}
\nc{\PRL}[3]{Phys.\ Rev.\ Lett.\ \textbf{#1}, #2 (#3)}
\nc{\PRA}[3]{Phys.\ Rev.\ A \textbf{#1}, #2 (#3)}
\nc{\PRB}[3]{Phys.\ Rev.\ B \textbf{#1}, #2 (#3)}
\nc{\PRC}[3]{Phys.\ Rev.\ C \textbf{#1}, #2 (#3)}
\nc{\PRD}[3]{Phys.\ Rev.\ D \textbf{#1}, #2 (#3)}
\nc{\PRE}[3]{Phys.\ Rev.\ E \textbf{#1}, #2 (#3)}
\nc{\PREP}[3]{Phys.\ Rep.\ \textbf{#1}, #2 (#3)}
\nc{\RMP}[3]{Rev.\ Mod.\ Phys.\ \textbf{#1}, #2 (#3)}
\nc{\Sc}[3]{Science \textbf{#1}, #2 (#3)}
\nc{\SPJETP}[3]{Sov.\ Phys.\ JETP \textbf{#1}, #2 (#3)}
\nc{\TMF}[3]{Teor.\ Mat.\ Fiz.\ \textbf{#1}, #2 (#3)}
\nc{\ZN}[3]{Z.\ Naturforsch.\ \textbf{#1}, #2 (#3)}
\nc{\ZPB}[3]{Z.\ Phys.\ B Condens.\ Matter \textbf{#1}, #2 (#3)}

\begin{document}

\title{Charge Transport and Quantum Phase Transitions in
Singlet Superconductor - Ferromagnet - Singlet Superconductor Junctions}
\author{Boris Kastening$^{1,2}$, Dirk K.~Morr$^{3,4}$,
Lambert Alff$^2$, and Karl Bennemann$^4$} \affiliation{ $^1$
Institut f\"ur Theoretische Physik, Technische Hochschule Aachen,
Physikzentrum, 52056 Aachen, Germany\\
$^2$ Institut f\"ur Materialwissenschaft, Technische Universit\"at
Darmstadt, Petersenstra{\ss}e 23, 64287 Darmstadt, Germany\\
$^3$ Department of Physics, University of Illinois at Chicago, Chicago,
Illinois 60607, USA\\
$^4$ Institut f\"ur Theoretische Physik, Freie Universit\"{a}t Berlin,
Arnimallee 14, 14195 Berlin, Germany}
\date{\today}
\begin{abstract}
We study the Josephson current, $I_J$, in a junction consisting of
two $s$-wave superconductors that are separated by a ferromagnetic
barrier possessing a magnetic and non-magnetic scattering potential,
$g$ and $Z$, respectively. We discuss the general dependence of
$I_J$ on $g$, $Z$, and the phase difference $\phi$ between the two
superconductors. Moreover, we compute the critical current, $I_c$
for given $g$ and $Z$, and show that it possesses two lines of
non-analyticity in the $(g, Z)$-plane. We identify those regions in
the $(g, Z)$-plane where the Josephson current changes sign with
increasing temperature without a change in the relative phase
between the two superconductors, i.e., without a transition between
a $0$ and $\pi$ state of the junction. Finally, we show that by
changing the relative phase $\phi$, it is possible to tune the
junction through a first-order quantum phase transition in which the
spin polarization of the two superconductors' combined ground state
changes from $\langle S_z \rangle =0$ to $\langle S_z \rangle =1/2$.

\end{abstract}


\maketitle

\section{Introduction}

Heterostructures consisting of magnetically active layers provide
new possibilities for manipulating charge (and potentially spin)
transport and are hence of great interest for the field of spin
electronics \cite{Wolf01}. Josephson junctions consisting of
conventional $s$-wave superconductors and a ferromagnetic barrier
fall into this category, and their study has led to the discovery of
a number of fundamentally new phenomena (for a recent review, see
Ref.~\onlinecite{reviews} and references therein). Among these is the
transition from a $0$-state to a
$\pi$-state in junctions with a metallic ferromagnetic barrier, which is
accompanied by a sign change (and hence directional change) of the Josephson
current. This transition signifies an intrinsic phase change of
$\pi$ between the superconductors forming the Josephson junction
which arises from a temperature dependent decay length and
oscillation length of the superconducting order parameter inside the
ferromagnetic metal. Such a transition was predicted theoretically
\cite{Bul77}, and subsequently also observed experimentally
\cite{Rya01}. It was recently shown, by using a
phenomenological $S$-matrix scattering formalism, that a sign change
of the Josephson current with increasing temperature can also occur
in insulating ferromagnetic barriers \cite{Fog00,Bar02}. Changing
the direction of the Josephson current by increasing temperature or
by varying the relative phase between the superconductors, leads to
novel types of current switches that possess potential applications
in quantum information technology \cite{Wolf01,Iof99}. Since most
ferromagnetic barriers do not only possess a magnetic scattering
potential, but also a non-magnetic one, the question naturally
arises to what extent the interplay between these two types of
scattering potentials either alters the effects discussed above or
leads to qualitatively new phenomena.

In this article, we study the Josephson current, $I_J$, in a
one-dimensional (1D) Josephson junction consisting of two $s$-wave
superconductors and a thin ($\delta$-function type) ferromagnetic
barrier (SFS junction).
We start from a microscopic Hamiltonian in which the
barrier possesses a magnetic and non-magnetic scattering potential,
described by $g$ and $Z$, respectively, the former being directly
proportional to the barrier's magnetization. We discuss the general
dependence of the charge Josephson current on $g$, $Z$, and the
relative phase, $\phi$, between the two superconductors.  In
particular, we demonstrate that in certain regions of the $(g,
Z)$-plane, $I_J$ varies continuously with $\phi$, while in other
regions, and particularly around $Z=g$, $I_J$ exhibits
discontinuities. We compute the critical current, $I_c$, defined as
the maximum Josephson current for a given $g$ and $Z$, and show that
it possesses two lines of non-analytic behavior in the $(g,
Z)$-plane. These non-analyticities correspond to discontinuities in
the first and second derivative of $I_c$ (with respect to $g$ or
$Z$). We show that $I_c$ exhibits qualitatively different
dependencies on the scattering strength in different parts of the
$(g, Z)$-plane, which opens the interesting (and quite
counterintuitive) possibility to increase the critical current
through the junction by increasing the junction's magnetization.
Moreover, we identify those regions of the $(g, Z)$-plane in which
the Josephson current changes sign (and thus direction) with
increasing temperature without a change in the relative phase
between the two superconductors, i.e., without a transition between
a $0$ and $\pi$ state of the junction. In addition, we find that
while the total spin Josephson current, $I_s$, flowing through the
junction is zero, there are two contributions to $I_s$, arising from
the Andreev and continuum states, respectively, that are equal in
magnitude but possess opposite signs. We show that if these two
contributions can be independently measured, this would open new
venues for employing the combined spin and charge degrees of freedom
in such a junction. Finally, we demonstrate that by changing the
phase $\phi$ between the superconductors, it is possible to tune the
junction through a first-order quantum phase transition in which the
spin polarization of the superconductors' ground state changes from
$\langle S_z \rangle =0$ to $\langle S_z \rangle =1/2$.

The theoretical methods used in this study provide direct insight
into the explicit dependence of the Josephson current and the
Andreev states on the magnetization of the junction, and into the
interplay between magnetic and non-magnetic scattering potentials.

\section{Theoretical Methods}

We take the 1D SFS junction to be aligned along the $z$-axis and to
be described by the Hamiltonian
\begin{widetext}
\beq \label{H}
{\cal H}=\int dz\left\{\sum_\si\psi^\dg_\si(z)
\left[-\f{\hbar^2\p_z^2}{2m}-\mu+U(z)\right]\psi_\si(z)
-\left[\De(z)\psi^\dg_\ua (z)\psi^\dg_\da(z)+\text{H.c.}\right]
-\frac{g_e \mu_B\mu_0}{\hbar}{\bf M}(z)\cdot
\sum_{\al,\be}\psi^\dg_\al(z)\vec{\bm{\si}}_{\al\be}\psi_\be(z)\right\}\ ,
\eeq
\end{widetext}
where $\psi^\dg_\si(z)$ and $\psi_\si(z)$ are the fermionic operators that
create or annihilate a particle with spin $\si$ at site $z$,
respectively. $\De(z)$ is the $s$-wave superconducting gap, and
$U(z)=U_0\de(z)$ describes the non-magnetic (i.e., potential)
scattering strength of the junction at $z=0$. Without loss of
generality, we choose the magnetization of the junction, ${\bf
M}(z)$, to be parallel to the $z$-axis, i.e., ${\bf
M}(z)=M_0(0,0,1)\de(z)$. In order to simplify the notation, we set
$g_e \mu_B\mu_0/\hbar=1$. Moreover, to facilitate the discussion of
the spin structure of the Josephson current and the junction's
ground state, we choose the quantization direction to coincide with
the direction of the magnetic moment, i.e., with the $z$-direction.

In the following, we use two complementary theoretical approaches in
order to compute the Josephson current through the interface: one
follows the Blonder-Tinkham-Klapwijk (BTK) approach \cite{BlTiKl82}
(see Sec.~\ref{BDGmethod}), the other one starts from the quantum
mechanical definition of the current operator and computes its
expectation value (see Sec.~\ref{QMmethod}).

\subsection{BTK Ansatz}
\label{BDGmethod}

At the interface between the two superconductors, two Andreev bound
states \cite{And64} with energies $E_{\al,\be}$ are formed
\cite{Bar02,Zik99}. As was discussed before \cite{BlTiKl82,Rie98,Zag98},
and as we will explicitly show in Sec.~\ref{QMmethod}, the charge Josephson
current flows solely through these two bound states, and is hence
given by \cite{Zag98}
\beq \label{IJ}
I_J=I_J^\al+I_J^\be
=-\f{e}{\hbar}\sum_{j=\al,\be}\f{\p E_j}{\p \phi}
\tanh\left(\f{E_j}{2 k_B T}\right)\ ,
\eeq
where $\phi$ is the phase
difference between the superconducting order parameters on the left
and right side of the junction. In order to compute $E_{\al,\be}$,
we start from Eq.~(\ref{H}) and derive the Bogoliubov-de Gennes
(BdG) equation \cite{Zag98,Fur99,Kwon04,Gen89} by introducing the
unitary Bogoliubov transformation
\bse \label{BdGtrafo}
\begin{align}
\psi_\ua(z) &= \sum_n u_{n,\alpha}(z) \alpha_{n}+v_{n,\beta}^\ast(z)
\beta_{n}^\dg\ , \label{eq:BT1}
\\
\psi_\da (z) &= \sum_n -u_{n,\beta}(z)
\beta_{n}+v_{n,\alpha}^\ast(z)\alpha_{n}^\dg\ , \label{eq:BT2}
\end{align}
\ese
where the sum runs over all eigenstates of the junction, and
$\alpha_{n}, \beta_{n}$ are quasi-particle operators in terms of
which the Hamiltonian, Eq.~(\ref{H}), is diagonal. Defining
\beq
\label{Psi}
\Psi_{n,j}(z)\equiv\bpm u_{n,j}(z)\\v_{n,j}(z)\epm,
\eeq
with $j={\al,\be}$, the BdG equation is given by
\beq \label{BdG}
\hat{H}_j\Psi_{n,j}(z)=E_{n,j}\Psi_{n,j}(z)\ .
\eeq
Here $E_{n,j}$ is the energy of the state $\Psi_{n,j}$,
\begin{equation}
\hat{H}_j=\left(\ba{cc}
H_0{\mp}H_M&-\De\\-\De^\ast&-H_0{\mp}H_M\ea\right)\ ,
\end{equation} where the upper (lower) sign corresponds to
$j=\alpha(\beta)$ and
\bse
\begin{align}
H_0\equiv&-\f{\hbar^2\p_z^2}{2m}-\mu+U_0\de(z) \ , \\
H_M\equiv& M_0\de(z) \ , \\
\De\equiv&
\begin{cases}
\De_0 & z < 0\ , \\
\De_0 e^{-i \phi} & z>0\ ,
\end{cases}
\end{align}
\ese
with $\De_0$ chosen real. The case $n=0$ corresponds to the
Andreev bound states, and in what follows, we use $E_j$ as a
shorthand notation for their energies, $E_{0,j}$.

For the bound states' wave-function on the left and right hand sides
of the junction, $\Psi_{0,j,L}(z)$ and $\Psi_{0,j,R}(z)$,
respectively, we make the ansatz
\beq
\label{ansatz}
\Psi_{0,j,s}(z)=e^{c_s \kappa_j z} \sum_{\de=\pm} \bpm
u_{0,j,s,\de}\\v_{0,j,s,\de}\epm e^{\de i k_F z}\ ,
\eeq
with $s=L,R$, $c_{L,R}=\pm 1$, and $k_F$ is the Fermi momentum. Note that
the decay length of the Andreev state,
$\kappa_j^{-1}=\hbar v_F/\sqrt{\Dz^2-E_j^2}$ with $v_F=\hbar k_F/m$, itself
depends on $E_j$. The solutions of Eq.~(\ref{BdG}) are subject to
the boundary conditions
\bse \label{BC}
\begin{align}
&
\Psi_{0,j,L}(0)=\Psi_{0,j,R}(0)\ ,
\\
\p_z\Psi_{0,j,R}(0)-\p_z
&
\Psi_{0,j,L}(0)=
\nn\\
\f{2m}{\hbar^2}
&
\left(\ba{cc}U_0{\mp}M_0 & \\ & U_0{\pm}M_0\ea\right)
\Psi_{0,j,R}(0)\ ,
\end{align}
\ese
where again the upper (lower) sign corresponds to $j=\alpha
(\beta)$. In the limit $k_F \gg \kappa_j$, which holds for
superconductors with coherence length $\xi_c=v_F/\Dz \gg 1/k_F$, the
solution of the BdG equation yields two Andreev states with energies
\begin{equation} \label{eab}
\f{E_j}{\Dz} =\f{\sqrt{A+B}\mp\sqrt{A-B}}
{\sqrt{2\left[ 1+(g+Z)^2 \right]\left[1+(g-Z)^2 \right]}}\ ,
\end{equation}
where the upper (lower) sign corresponds to $j=\alpha (\beta)$
and
\bse \label{ab}
\begin{align}
A&=(1+\Z^2-g^2)\left[\cos^2\left(\phi/2 \right)+\Z^2-g^2\right]+2g^2\ ,\\
B&=\sqrt{\left[ 1+(g+Z)^2 \right]\left[1+(g-Z)^2 \right]}
\nn\\
&\phantom{=}\times\left[\cos^2\left(\phi/2 \right)+\Z^2-g^2\right]\ ,
\end{align}
\ese
with $g=m M_0/\hbar^2k_F$ and $\Z=mU_0/\hbar^2k_F$. Without loss of
generality we assume $g,\Z\geq0$ from here on. While $E_\be$
does not change sign as a function of $\phi$ (for $Z \not = 0$),
$E_\alpha$ changes sign if $0<g^2-Z^2<1$ (hence for $g \leq Z$
or $g\geq \sqrt{1+Z^2}$, no sign change of either bound state takes
place). This sign change, which occurs at a phase difference
$\phi_c^\alpha$ given by $\cos^2(\phi_c^\alpha/2)=g^2-Z^2$,
indicates a first-order phase transition in which the spin
polarization of the superconductors' ground state changes, as
discussed in more detail in Sec.~\ref{sec:firstorder}.

For the subsequent discussion, it is necessary to consider the spin
structure of the Andreev states. To this end, we compute the local
density of states (LDOS), $N(\sigma,z)$, (i.e., the local spectral
function) for the \mbox{spin-$\ua$} and \mbox{spin-$\da$} components
of the Andreev states, which, using the Bogoliubov transformation
presented in Eq.~(\ref{BdGtrafo}), are readily obtained as
\bse
\label{eq:Ak}
\begin{align}
N(\ua,z)&=|u_{0,\alpha}(z)|^2
\delta(\omega{-}E_\alpha)+|v_{0,\beta}(z)|^2 \delta(\omega{+}E_\beta)\ ,
\label{eq:Ak1} \\
N(\da,z)&=|u_{0,\beta}(z)|^2
\delta(\omega{-}E_\beta)+|v_{0,\alpha}(z)|^2
\delta(\omega{+}E_\alpha)\ . \label{eq:Ak2}
\end{align}
\ese
Hence, (for $E_{\alpha (\beta)}>0$) the wave-function of the
$\alpha$ state, $|\Psi_{0,\alpha} \rangle$, possesses a
particle-like \mbox{spin-$\da$} component, and a hole-like
\mbox{spin-$\ua$} component, i.e., $|\Psi_{0,\alpha} \rangle=|p,\da
\rangle + |h,\ua\rangle$. Similarly for the $\beta$-state,
$|\Psi_{0,\beta} \rangle= |p,\ua \rangle + |h,\da \rangle$. When
$E_\alpha$ changes sign, the occupation numbers of the
\mbox{spin-$\ua$} and \mbox{spin-$\da$} components of the respective
wave-function are interchanged. We thus find that the LDOS near the
junction barrier contains four peaks inside the superconducting gap,
in agreement with the results obtained in
Refs.~\onlinecite{Bar02,Zik99}. As an important check of our
calculations, we consider the limit $\phi=0$, where the Josephson
junction is identical to a system, in which a single magnetic
impurity is located inside a 1D s-wave superconductor. In this
limit, Eq.~(\ref{eab}) yields that only one of the Andreev states
exists inside the superconducting gap with $E_\alpha<\Delta_0$,
while the other Andreev state possesses the energy
$E_\beta=\Delta_0$ and is thus part of the continuum (see
Fig.~\ref{EandIj}(a)). These results are in agreement with those of
${\hat T}$-matrix and Bogoliubov-de Gennes approaches used in the
context of impurity scattering in $s$-wave superconductors (for a
recent review, see Ref.~\onlinecite{Balatsky_RMP}). We note that in
this limit, $\phi=0$, our results (and those of
Ref.~\onlinecite{Balatsky_RMP}) for $E_{\alpha,\beta}$
do not agree with the findings in Refs.~\onlinecite{Fog00,Bar02}
[see also Ref.~\onlinecite{com1}].

\subsection{Quantum Mechanical Current}
\label{QMmethod}

Using the operator definition of the quantum mechanical current, we
may resolve the spin-$\ua$ and spin-$\da$ particle currents
$I_\ua(z)$ and $I_\da(z)$, which in turn allows to compute the
charge and spin currents via
\bse \label{IJIS}
\begin{align}
\label{IJua}
I_J(z)&\equiv -e[I_\ua(z)+I_\da(z)],\\
\label{ISua}
I_S(z)&\equiv \f{\hbar}{2}[I_\ua(z)-I_\da(z)].
\end{align}
\ese

In order to obtain appropriately defined current operators
$\hat{I}_\ua(z)$ and $\hat{I}_\da(z)$, whose expectation values are
the currents $I_\ua(z)$ and $I_\da(z)$, we note that the density
operator of spin-$\si$ electrons is
\beq \label{eq:rhodef1}
\hat{\rho}_\si(z) =\left.\hat{\rho}_\si(z,z')\right|_{z'=z},
\eeq
where
\begin{align}
\label{rhodef}
\hat{\rho}_\si(z,z') &\equiv
\psi_\si^\dg(z)\psi_\si(z').
\end{align}
The quantum mechanical particle current operator corresponding to
$\hat{\rho}_\si(z)$ then follows from
\begin{align}
\label{Isigmatot}
\hat{I}_\si(z)
&=
\left.\f{i\hbar}{2m}\left[(\p_z-\p_{z'})
\hat{\rho}_\si(z,z')\right]\right|_{z'=z}
\nn\\
&=
\f{i\hbar}{2m}
\left\{[\p_z\psi_\si^\dg(z)]\psi_\si(z)
-\psi_\si^\dg(z)[\p_z\psi_\si(z)]\right\},
\end{align}
and its expectation value
\begin{align}
\label{eq:Isiasym} I_\si(z) &= \f{\hbar}{m}\text{Im}
\left\langle\psi_\si^\dg(z)[\p_z\psi_\si(z)] \right\rangle
\end{align}
is the corresponding particle current. In what follows, we refer to
$I_\si(z)$ as the ``conventional" form of the particle current.

After diagonalizing the Hamiltonian with the Bogoliubov
transformation of Eq.~(\ref{BdGtrafo}), the currents given in
Eq.~(\ref{IJIS}) possess in general contributions from both the
Andreev bound states and the continuum states. The calculation of
the latter is prohibitively cumbersome when using the form of
$I_\sigma$ given in Eq.~(\ref{eq:Isiasym}). It turns out, however,
that one can use an alternative formulation to evaluate $I_\si$ by
defining a ``symmetrized" form of the current operator.
Specifically, using the anticommutator
\begin{align}
\label{anticom}
\{\psi_\si^\dg(z),\psi_{\si'}(z')\}&=\de_{\si\si'}\de(z-z'),
\end{align}
we can write
\begin{align}
\label{eq:rho1} \hat{\rho}_\si(z,z') &\equiv \f{1}{2} \left[
\de(z-z') + \psi_\si^\dg(z)\psi_\si(z')
-\psi_\si(z')\psi_\si^\dg(z)\right].
\end{align}
We next define a ``symmetrized" density via
\begin{align}
\label{rhosymdef} \hat{\rho}^\text{sym}_\si(z,z') &\equiv
\f{1}{2}\left[\psi_\si^\dg(z)\psi_\si(z')
+\psi_\si(z)\psi_\si^\dg(z')\right] \ ,
\end{align}
where the second term on the r.h.s.\ of Eq.~(\ref{rhosymdef}) arises
from the first term via a particle-hole transformation (this fact
becomes important when discussing the form of the LDOS corresponding
to $\hat{\rho}^\text{sym}_\si(z,z)$, see below). Since the
$\de$-function in Eq.~(\ref{eq:rho1}) does not contribute to the
particle current operator, it immediately follows that the densities
defined in Eqs.~(\ref{eq:rho1}) and (\ref{rhosymdef}) when inserted
into Eq.~(\ref{Isigmatot}) yield the same spin-$\si$ particle current
operator. We can therefore write $\hat I_{\si}$ in a ``symmetrized"
form as
\begin{align}
\label{Isigma} \lefteqn{\hat I_{\si}(z) =
\left.\f{i\hbar}{2m}
\left[(\p_z-\p_{z'})\hat{\rho}^\text{sym}_\si(z,z')\right]\right|_{z'=z}}
\nn\\
&= -\f{i\hbar}{4m} \left\{ \psi_\si^\dg(z)[\p_z\psi_\si(z)]
+\psi_\si(z)[\p_z\psi_\si^\dg(z)] -\text{h.c.}\right\} \ .
\end{align}
One then obtains
\begin{align}
\label{eq:IsSym} I_{\si}(z) &= \f{\hbar}{2m}\text{Im}
\left\langle\psi_\si^\dg(z)\p_z\psi_\si(z)
+\psi_\si(z)\p_z\psi_\si^\dg(z)\right\rangle
\end{align}
as the corresponding ``symmetrized" form of the spin-$\si$ particle
current. While the current thus defined is identical to the
conventional form given in Eq.~(\ref{eq:Isiasym}), it turns out that
the calculation of the contributions from continuum states using the
right hand side of Eq.~(\ref{eq:IsSym}) is considerably simplified.
In what follows, we therefore discuss the contributions of continuum
and Andreev states to the charge and spin current using the
symmetrized from of the particle current, Eq.~(\ref{eq:IsSym}). Since
the contributions to the particle current arising from individual
(Andreev or continuum) states are different in the conventional and
the symmetrized form, we distinguish between them by denoting with
(without) a tilde the current flowing through a specific state
within the symmetrized (conventional) definition. We then obtain
\begin{align}
\label{eq:Iudtotal}
I_{\ua(\da)} (z) &= \sum_n  \tilde{I}_{\ua(\da),n} (z),
\end{align}
where
\bse
\label{eq:Iudsingle}
\begin{align}
\label{Iu}
\tilde{I}_{\ua,n}(z)
&\equiv
I_{n,\al}^u(z)\tanh\f{\be E_{n,\al}}{2}
+I_{n,\be}^v(z)\tanh\f{\be E_{n,\be}}{2},
\\
\tilde{I}_{\da,n}(z)
&\equiv
I_{n,\be}^u(z)\tanh\f{\be E_{n,\be}}{2}
+I_{n,\al}^v(z)\tanh\f{\be E_{n,\al}}{2},
\end{align}
\ese
with amplitude functions
\begin{align}
I^f_{n,j}(z) &\equiv
\f{\hbar}{2m}\text{Im}[\ff_{n,j}(z)\p_z\fast_{n,j}(z)],
\end{align}
where $f=u,v$ and $j=\al,\be$, and the sum in Eq.~(\ref{eq:Iudtotal})
runs over all states of the system (including
Andreev and continuum states).

We begin by computing the contribution to the total current arising
from the continuum states. Using the general ansatz for the form of
these states described in Appendix \ref{sec:CSdiag}, we find that
the continuum states with momentum $\pm(k_F \pm q)$ are degenerate
and that their energy $E_q$ is given by (in the limit $E_q \ll \mu$)
\beq \label{E2supergapS}
E_q^2=|\Dz|^2+\left(\f{\hbar^2k_Fq}{m}\right)^2.
\eeq
Diagonalizing
the corresponding subspace, one obtains three important relations
\bse \label{IuapmIda}
\begin{align}
\label{IuapIda} & \sum_{n(E_q)}[{\tilde I}_{\ua,n}(z)+{\tilde
I}_{\da,n}(z)] \propto\sin2q|z|,
\\
\label{IuamIda} & \sum_{n(E_q)}[{\tilde I}_{\ua,n}(z)-{\tilde
I}_{\da,n}(z)] =0, \\
\label{IuIv} & \sum_{n(E_q)} I^u_{n,\alpha (\beta)}(z=0)
=\sum_{n(E_q)} I^v_{n,\alpha (\beta)}(z=0)\ ,
\end{align}
\ese
where the sums run over an orthonormal basis of continuum
states with energy $E_q$. The first relation, Eq.~(\ref{IuapIda}),
implies that given the definition of the charge Josephson current,
$I_J$, in Eq.~(\ref{IJua}), the contribution to $I_J$ from continuum
states at the interface is identically zero within the symmetrized
version of the current. As we show below, this result also holds
when the conventional definition of the current operator is used.
General arguments have been put forward that this result arises
since the density of continuum states in the presence of a barrier
is unchanged \cite{Zag98Sec453}. However, away from the interface the
continuum states carry a non-zero charge current since charge
conservation requires that the (decaying) current through the Andreev
states be compensated by a current carried through the continuum states.
The second relation, Eq.~(\ref{IuamIda}), implies that the contribution
of the continuum states to the spin Josephson current, as defined in
Eq.~(\ref{ISua}), is zero at any position $z$ along the junction.
Finally, the third relation, Eq.~(\ref{IuIv}), when combined with
Eq.~(\ref{eq:Iudsingle}), yields that at the barrier, there are no
contributions from the continuum states to either the spin-$\ua$ or
spin-$\da$ particle current.

Since the continuum states carry no charge current at the barrier
(i.e., at $z=0$) the total charge Josephson current is solely carried
by the Andreev states and thus given by
\beq
\label{IJIJ0} I_J\equiv I_J(0)=-e[{\tilde
I}_\ua^\text{AS}(0)+{\tilde I}_\da^\text{AS}(0)] \ ,
\eeq
where ${\tilde I}_{\ua,\da}^\text{AS}$ are the currents through the
Andreev states in the symmetric formulation of the current. These are
given by the $n=0$ term in Eq.~(\ref{eq:Iudtotal}) for which one thus has
\bse \label{Iudsym}
\begin{align}
\label{IuASsym} {\tilde I}_\ua^\text{AS}(z) &=
I_{0,\al}^u(z)\tanh\f{\be E_\al}{2} +I_{0,\be}^v(z)\tanh\f{\be
E_\be}{2},
\\
\label{IdASsym} {\tilde I}_\da^\text{AS}(z) &=
I_{0,\be}^u(z)\tanh\f{\be E_\be}{2} +I_{0,\al}^v(z)\tanh\f{\be
E_\al}{2}.
\end{align}
\ese It is straightforward to show that the bound state ansatz of
Eq.~(\ref{ansatz}) leads to
\bse \label{IualIube}
\begin{align}
\label{eq:Iuva} I_{0,\al}^u(z)&=I_{0,\al}^v(z)\propto e^{-2\ka|z|},
\\
\label{eq:Iuvb} I_{0,\be}^u(z)&=I_{0,\be}^v(z)\propto e^{-2\ka|z|},
\end{align}
\ese
yielding ${\tilde I}_\ua^\text{AS}(z)={\tilde
I}_\da^\text{AS}(z)$. Together with Eq.~(\ref{IuamIda}), this result
implies that within the symmetrized form of the particle current,
Eq.~(\ref{eq:IsSym}), neither the continuum states nor the Andreev
bound states carry a spin current. Hence, we obtain that the total
spin current $I_S(z)$ defined in Eq.~(\ref{ISua}) vanishes at any
point along the junction, in agreement with the arguments in
Ref.\cite{Mic08}. This result holds even when the system undergoes a
first order quantum phase transition in which the spin polarization
of the junction's ground state changes (see
Sec.~\ref{sec:firstorder}).

While the symmetrized form of the Josephson current used above
allowed for a simpler evaluation of the contributions arising from
the continuum states, any physical interpretation of the Josephson
current has to be based on its conventional definition, given by
Eq.~(\ref{Isigmatot}), and the form of the spin density in
Eq.~(\ref{rhodef}). In what follows, we therefore discuss the form of
the charge and spin currents within the conventional definition, and
compare them with the symmetrized results presented above. A
connection between the expressions for $\tilde{I}_{\sigma,n}$ and
$I_{\sigma,n}$ (which represent the currents flowing through state
$n$ in the symmetrized and conventional definition, respectively)
can be made by using the following identities
\bse
\label{Iidsalbepap}
\begin{align}
\sum_n[I_{n,\al}^u(z)-I_{n,\be}^v(z)]&=0, \label{eq:Id1} \\
\sum_n[I_{n,\be}^u(z)-I_{n,\al}^v(z)]&=0, \label{eq:Id2}
\end{align}
\ese
where, as in Eq.~(\ref{eq:Iudtotal}), the sum runs over all
states of the junction. These identities are derived by applying
$\p_z-\p_{z'}$ to the anticommutator in Eq.~(\ref{anticom}), and
subsequently setting $\si'=\si$ and $z'=z$, and using the form of
$\psi_\ua(z)$ and $\psi_\da(z)$ given in Eq.~(\ref{BdGtrafo}). One
then finds that by subtracting the left hand side of
Eq.~(\ref{eq:Id1}) [Eq.~(\ref{eq:Id2})] from the right hand side of
Eq.~(\ref{eq:Iudtotal}) for $\sigma=\ua$ [$\sigma=\da$], one obtains
the corresponding expressions for $I_{\sigma,n}$. In particular,
within the conventional definition, the Josephson current through
the Andreev bound states is given by
\bse \label{eq:IAS}
\begin{align}
\label{IuAS} I_\ua^\text{AS}(z) &=
-I_{0,\al}^u(z)\left(1-\tanh\f{\be E_\al}{2}\right)\nn\\
&\phantom{=} +I_{0,\be}^v(z)\left(1+\tanh\f{\be
E_\be}{2}\right),
\\
\label{IdAS} I_\da^\text{AS}(z) &=
-I_{0,\be}^u(z)\left(1-\tanh\f{\be E_\be}{2}\right)\nn\\
&\phantom{=} +I_{0,\al}^v(z)\left(1+\tanh\f{\be
E_\al}{2}\right).
\end{align}
\ese
Given the result for $I_J$ in Eq.~(\ref{IJIJ0}), it immediately
follows that
\begin{equation}
I_J=-e\left[I_\ua^\text{AS}(0)+I_\da^\text{AS}(0)\right] \ ,
\end{equation}
implying that also within the conventional definition of the
particle current, the total charge Josephson current at $z=0$ is
solely carried by the Andreev states with no contribution arising
from the continuum states. This result justifies the use of
Eq.~(\ref{IJ}) in the BTK approach of Sec.~\ref{BDGmethod} for the
calculation of the total charge Josephson current. Moreover, since
the charge Josephson currents computed within the BdG approach of
Sec.~\ref{BDGmethod} should be the same as that of the quantum
mechanical method of Sec.~\ref{QMmethod}, one requires [by combining
Eqs.~(\ref{IJ}), (\ref{IJIJ0}), (\ref{Iudsym}), and
(\ref{IualIube})] that the following important identity be satisfied
\begin{align}
\label{eq:Identity} \f{\p E_j}{\p\phi}&=2\hbar I^v_{0j}(0),
\end{align}
with $j=\al,\be$. We have carried out an extensive numerical survey
in the parameter space of our system and found this relation to
always hold.

We next consider the form of the spin current in the conventional
definition of the currents. While the result of a vanishing total
spin current, that we obtained within the symmetrized form, also has
to hold within the conventional framework, we find that the
contributions from the Andreev and continuum states differ. In
particular, using the results of Eq.~(\ref{eq:IAS}), we find that the
contribution of the Andreev states to the spin current, $I^{AS}_S$,
is given by
\begin{equation}
I^{AS}_S(z)=\hbar \left[ I_{0,\be}^v(z)-I_{0,\al}^v(z) \right],
\end{equation}
which in general does not vanish. However, since the total spin
current still needs to be zero, it immediately follows that within
the conventional definition of the particle current, a spin current
of equal magnitude, but opposite sign to $I^{AS}_S$ flows through
the continuum states. Note, however, that while a spin Josephson
current flows through the continuum states, the charge current
through these states is zero, as discussed above. The contribution
to the spin current provided by the continuum states thus
compensates the spin current through the Andreev states and leads to
a zero total spin current [we return to a discussion of these two
(opposite) contributions to the spin current in
Sec.~\ref{sec:spintransport}]. In contrast, in the symmetrized
version, the spin current through each of the continuum states and the
Andreev states is exactly zero.

In order to understand the different origin of the zero spin current
in these two frameworks, and to gain insight into its physically
correct interpretation, we consider the local density of states of
the Andreev states, $N^\text{sym}(\sigma,z)$, corresponding to the
density $\rho_\sigma^\text{sym}(z)=\langle
\hat{\rho}_\sigma^\text{sym}(z,z) \rangle$, which is given by
\bse
\label{eq:Aksym}
\begin{align}
N^\text{sym}(\ua,z)&=\frac{1}{2}|u_{0,\alpha}(z)|^2 \left[
\delta(\omega - E_\alpha)+\delta(\omega+E_\alpha) \right]
\nn \\
&+ \frac{1}{2} |v_{0,\beta}(z)|^2 \left[
\delta(\omega{+}E_\beta)+\delta(\omega{-}E_\beta) \right] \
, \label{eq:Ak1ph} \\
N^\text{sym}(\da,z)&=\frac{1}{2}|u_{0,\beta}(z)|^2 \left[
\delta(\omega{-}E_\beta) + \delta(\omega{+}E_\beta) \right]
\nn \\
& +\frac{1}{2}|v_{0,\alpha}(z)|^2 \left[
\delta(\omega{+}E_\alpha) + \delta(\omega{-}E_\alpha)
\right]\ . \label{eq:Ak2ph}
\end{align}
\ese
Here, the second term on the right hand side of each equation
is obtained from the first one via a particle-hole transformation,
in agreement with the definition  of
$\hat{\rho}^\text{sym}_\si(z,z')$ in Eq.~(\ref{rhosymdef}). As a
result, the system now possesses two sets of two degenerate bound
states, i.e., a total of four Andreev bound states, each with a
spectral weight of $1/2$. Within each set, the degenerate bound
states differ by their spin quantum number: one bound state
possesses a particle (hole) branch which is spin-$\ua$ (spin-$\da$),
and vice versa for the second bound state. Using the relations in
Eqs.~(\ref{IualIube}) and (\ref{eq:Identity}), and the definition of
the Josephson current in Eqs.~(\ref{IJ}) and (\ref{IJua}), one
immediately finds that such a LDOS leads to the expressions for
${\tilde I}_{\ua, \da}^\text{AS}(z)$ given in Eq.~(\ref{Iudsym}).
Since the spin quantum numbers are opposite between the degenerate
bound states, it naturally follows that the spin current through the
Andreev states is zero. Note, however, that the LDOS given in
Eq.~(\ref{eq:Aksym}) is unphysical: it does not reflect the
symmetry-breaking of the ferromagnetic barrier since for every
spin-$\ua$ branch, there exists a degenerate spin-$\da$ branch.
Moreover, in the limit $\phi=0$ (where the barrier represents a
single magnetic impurity in a 1D s-wave superconductor), the
symmetrized LDOS of Eq.~(\ref{eq:Aksym}) is in disagreement with that
obtained from ${\hat T}$-matrix and Bogoliubov-de Gennes approaches
\cite{Balatsky_RMP}. In contrast, the LDOS of Eq.~(\ref{eq:Ak}) which
is based on the conventional definition of the density operator
reflects the symmetry-breaking of the ferromagnetic barrier. It is
therefore physical and in full agreement with the results of
Ref.~\cite{Balatsky_RMP} for $\phi=0$. It then follows that the
physically correct interpretation regarding the origin of a zero
total spin current is that both the Andreev states and the
continuum states carry a spin current of equal magnitude, but
opposite sign. We propose that this conclusion can be tested by
using a spin-resolved STM experiment which can distinguish between
the LDOS presented in Eq.~(\ref{eq:Ak}) on one hand, and that given
in Eqs.~(\ref{eq:Aksym}) on the other hand.

\section{Results}

\subsection{Charge Transport}

%
% Figure 1
%
\begin{figure*}
\makebox[17.5cm]{
\includegraphics[width=15.0cm,angle=0]{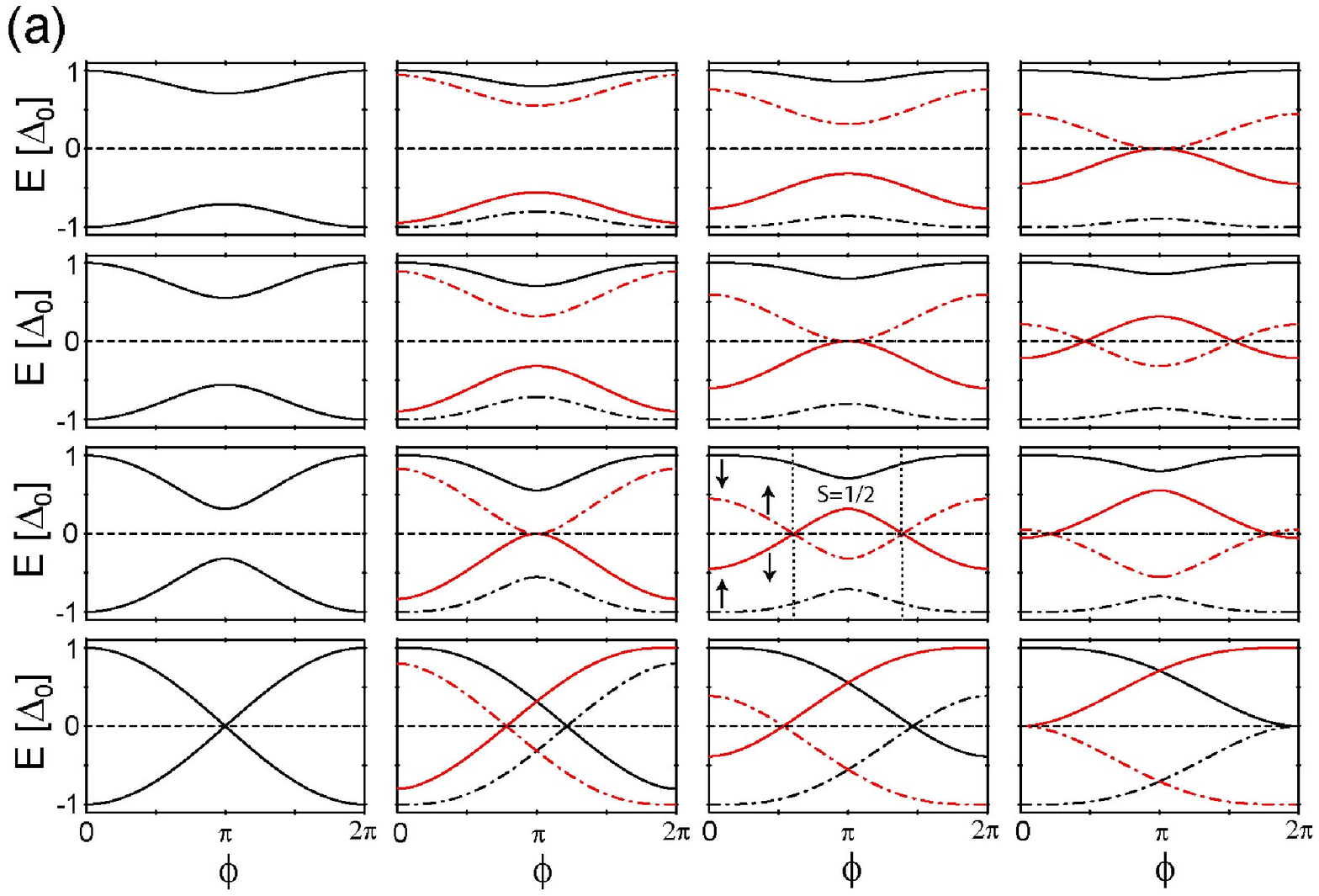}}
\makebox[17.5cm]{
\includegraphics[width=15.0cm,angle=0]{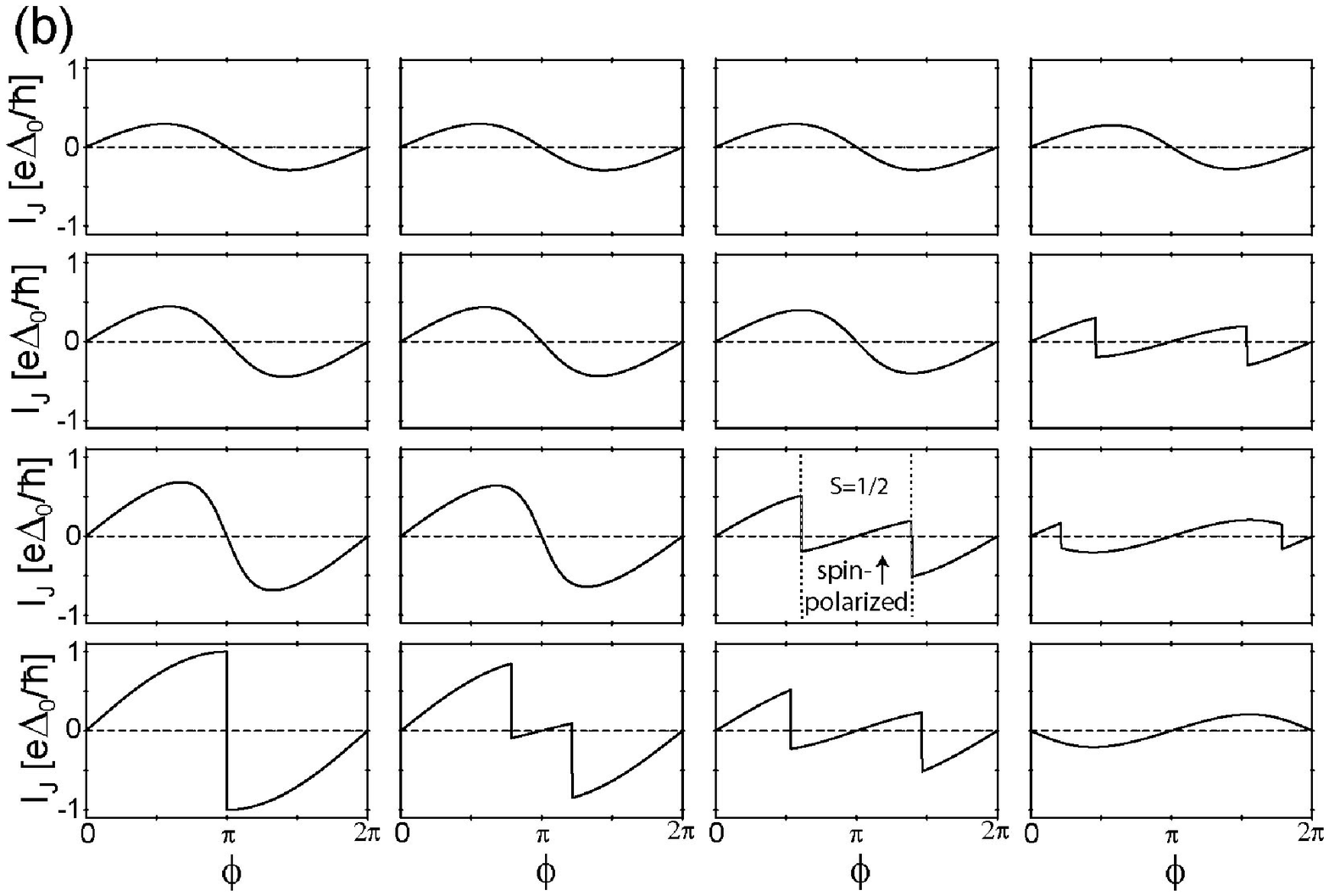}}
\caption{\label{EandIj} (Color online). (a) Energies $\pm
E_{\al,\be}$ of the Andreev states.
The columns from left to right correspond to $g=0,\f{1}{3},\f{2}{3},1$,
respectively, while the rows from bottom to top correspond to
$\Z=0,\f{1}{3},\f{2}{3},1$, respectively.
The energies of the $\alpha$ and
$\beta$ states are indicated by red (grey in print) and black lines,
respectively, while the \mbox{spin-$\ua$} and \mbox{spin-$\da$} components
are indicated by dashed-dotted and solid lines. For $g=0$ (left column)
the $\alpha$ and $\beta$ states are degenerate. (b) The resulting
$I_J$ at $T=0$ for several $g$ and $\Z$ as a function of $\phi$.}
\end{figure*}

The general dependence of the Andreev state energies and the
Josephson current on the scattering strength of the barrier is
presented in Fig.~\ref{EandIj} where we plot  $\pm E_{\al,\be}$
[Fig.~\ref{EandIj}(a)] and the resulting Josephson current at $T=0$
[Fig.~\ref{EandIj}(b)] as a function of $\phi$ for several values of
$Z$ and $g$ (the columns in Figs.~\ref{EandIj}(a) and
\ref{EandIj}(b) from left to right correspond to
$g=0,\f{1}{3},\f{2}{3},1$, respectively, while the rows from bottom
to top correspond to $\Z=0,\f{1}{3},\f{2}{3},1$, respectively.) For
a purely non-magnetic barrier [i.e., $g=0$, left column of
Fig.~\ref{EandIj}(a)], we find in agreement with earlier results
\cite{Bar02,Zag98,Fur99,Kwon04,zdef} that the Andreev states are
degenerate with energies
\begin{equation} \label{Enmag}
\f{E_{\al,\be}}{\Dz}=\sqrt{\f{\cos^2\left(\phi/2\right)+Z^2}{1+\Z^2}}\
.
\end{equation}
This degeneracy is lifted by  a non-zero magnetic scattering
potential of the junction (i.e., $g \not =0$) as shown in the three
right columns of Fig.~\ref{EandIj}(a). A qualitatively similar
result was also found in Refs.~\onlinecite{Bar02,Zik99,com1,Mic08}.
Specifically, for a purely magnetic scattering potential of the
junction [i.e., $Z=0$, bottom row of Fig.~\ref{EandIj}(a)], the
energies of the Andreev states are given by
\begin{equation}
 \label{eq:Emag}
\f{E_{\al,\be}}{\Dz}=\f{1}{1+g^2} \left[\cos\left(\phi/2 \right)\mp
g\sqrt{g^2+\sin^2\left(\phi/2 \right)}\right]\ .
\end{equation}
In agreement with the analytical results presented in
Eqs.~(\ref{Enmag}) and (\ref{eq:Emag}) we find that with increasing
$g \gg Z$ [plots in the lower right corner of Fig.~\ref{EandIj}(a)]
and $Z \gg g$ [plots in the upper left corner of
Fig.~\ref{EandIj}(a)] the energies of both bound states move toward
the gap edge. An interesting situation occurs for $g=Z$ [plots along
the diagonal of Fig.~\ref{EandIj}(a)], since in this case, the
effective scattering strength for the \mbox{spin-$\da$} and
\mbox{spin-$\ua$} electrons is $V^\da_\text{eff}=g+Z=2Z$ and
$V^\ua_\text{eff}=g-Z=0$,
respectively. In the unitary scattering limit $g=Z\gg1$, we then
find that the energies of the Andreev states are given by
\bse
\begin{align}
\f{E_{\alpha}}{\Dz}=&\f{\cos^2\left(\phi/2\right)}{2Z} +{\cal
O}(Z^{-3}) \label{eq:E_gz}\ ,\\
\f{E_{\beta}}{\Dz}=&1-\f{\sin^4\left(\phi/2\right)}{8Z^2} +{\cal
O}(Z^{-4})\ .
\end{align}
\ese
Hence, in the limit $g=Z\ra\infty$, the $\alpha$-state  becomes
a zero energy (midgap) state, while the $\beta$-state  moves into
the continuum. This analytical result is confirmed by the numerical
results shown in the plots along the diagonal of
Fig.~\ref{EandIj}(a).

We next discuss the form of the Josephson current, resulting from
the form of the Andreev states shown in Fig.~\ref{EandIj}(a). In the
unitary scattering limit $Z \gg \max\left\{g,1\right\}$ the
Josephson current at $T=0$ is given by (to leading order in $Z$)
\begin{equation}
\label{IJnmag} I_J=\f{e \Dz}{\hbar}\f{\sin\phi}{2Z^2} \ ,
\end{equation}
while for $g \gg \max\left\{Z,1\right\}$ one obtains to leading
order in $g$
\begin{equation} \label{IJmag} I_J=-\f{e \Dz}{\hbar}
\f{\sin\phi}{2g^2}\ .
\end{equation}
Hence, the Josephson currents for a predominantly non-magnetic
[Eq.~(\ref{IJnmag})] and predominantly magnetic [Eq.~(\ref{IJmag})]
barrier differ by a phase shift of $\pi$ in the unitary scattering
limit. This result also follows from a comparison of the $I_J$ plots
in the upper left ($Z \gg g$) and lower right ($g \gg Z$) corners of
Fig.~\ref{EandIj}(b).

Whether the Josephson current in the junction considered here is
carried by Cooper pairs, or by single electrons, depends on the
relative strength of $g$ and $Z$. We first recall that in a purely
non-magnetic junction (i.e., $g=0$) it was argued that the
dependence of the Josephson current on $Z$ in the unitary scattering
limit, $I_J \sim Z^{-2}$ [see Eq.~(\ref{IJnmag})], implies that the
current is carried by Cooper pairs \cite{Fur99,Kwon04}. In contrast,
in Josephson junctions consisting of unconventional superconductors
the scaling of the Josephson current, $I_J \sim Z^{-1}$, implies
that it is carried by single electrons \cite{Kwon04}. Here, we find
that for a predominantly magnetic junction with $g \gg \max\{Z,1\}$,
$I_J$ also scales with the inverse square of the scattering strength
(see Eq.~(\ref{IJmag}) and the current should thus also be carried
by Cooper pairs. In contrast, for the case $Z=g \rightarrow \infty$,
we obtain $I_J \sim Z^{-1}$ and the Josephson current should thus be
carried by single electrons. Further support for this conclusion
comes from considering the dependence of $E_\alpha$ on the
scattering strength in the limit $\phi=0$. As mentioned above, in
this case, the junction is identical to a static impurity in a (1D)
$s$-wave superconductor \cite{Shiba68}. If the impurity is purely
magnetic, the Andreev state $\alpha$ (which is better known in this
context as a Shiba state) is formed through scattering processes
involving the creation and destruction of Cooper-pairs. This
immediately follows from a diagrammatic derivation of the scattering
${\hat T}$-matrix which includes diagrams that contain the anomalous
Greens functions, $F$ and $F^*$, representing the creation and
destruction of Cooper-pairs, respectively \cite{Shiba68}. As a
direct result of the included anomalous Greens functions, one
obtains $E_\alpha \sim g^{-2}$, as in Eq.~(\ref{eq:Emag}) which
immediately leads to $I_J \sim g^{-2}$ for $\phi \not = 0$. In
contrast, for an impurity with scattering strength
$V^\da_\text{eff}=2Z$ and $V^\ua_\text{eff}=0$, the ${\hat
T}$-matrix is given by a series of diagrams that contain the normal
Greens function only (diagrams containing $F$ and $F^*$ are
forbidden), and hence $E_\alpha \sim Z^{-1}$. Note that for the same
reason, Josephson junctions consisting of unconventional
superconductors possess Andreev states whose energies also scale as
$E_i \sim Z^{-1}$. This connection between the scaling of $E_\alpha$
and $I_J$ for $\phi \not = 0$ with the case of impurity scattering
for $\phi = 0$ demonstrates that the presence or absence of
scattering diagrams involving the anomalous Greens function
determines the nature of the Josephson current.

\begin{figure}
\includegraphics[width=7.5cm,angle=0]{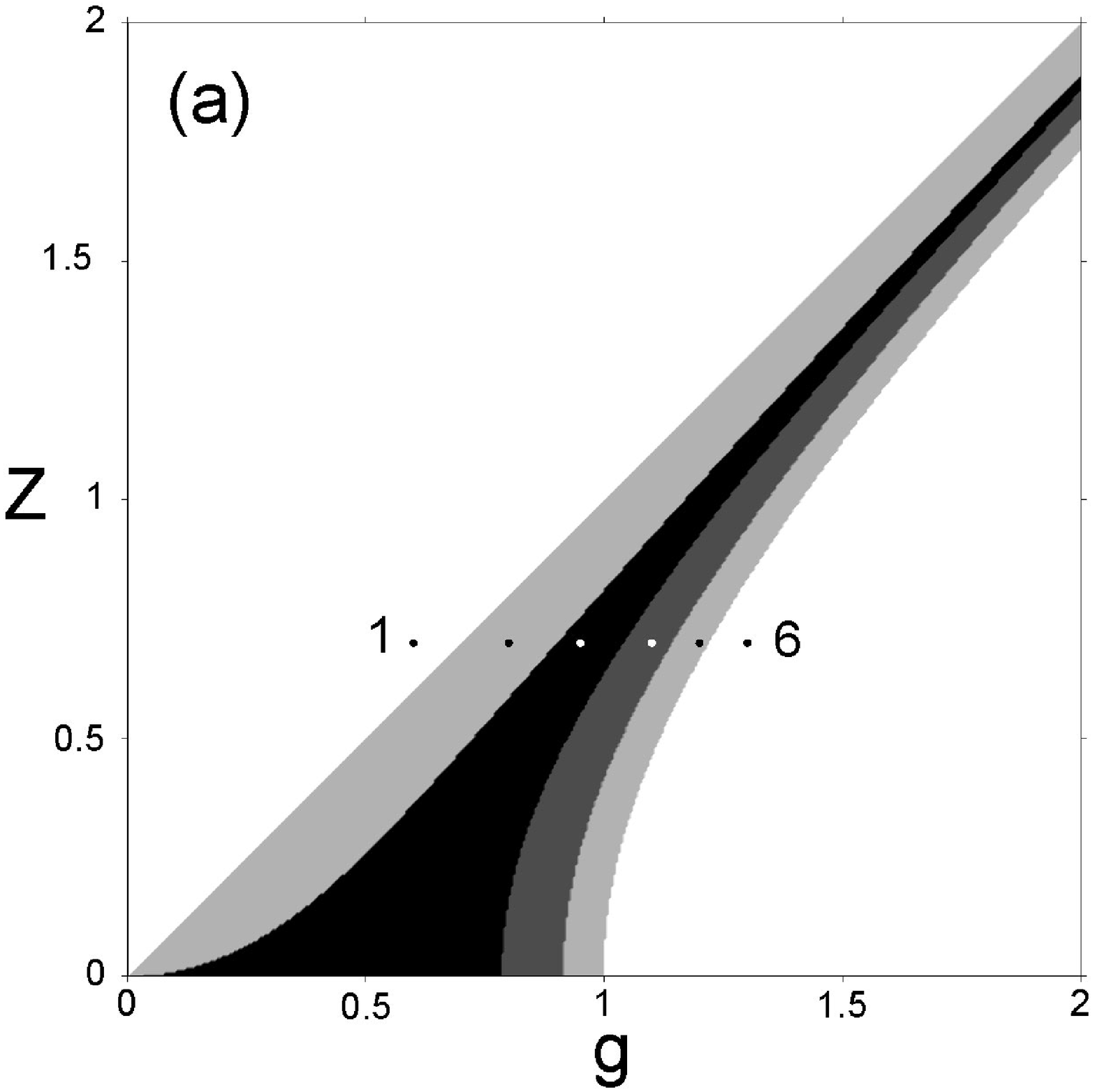}
\includegraphics[width=7.5cm,angle=0]{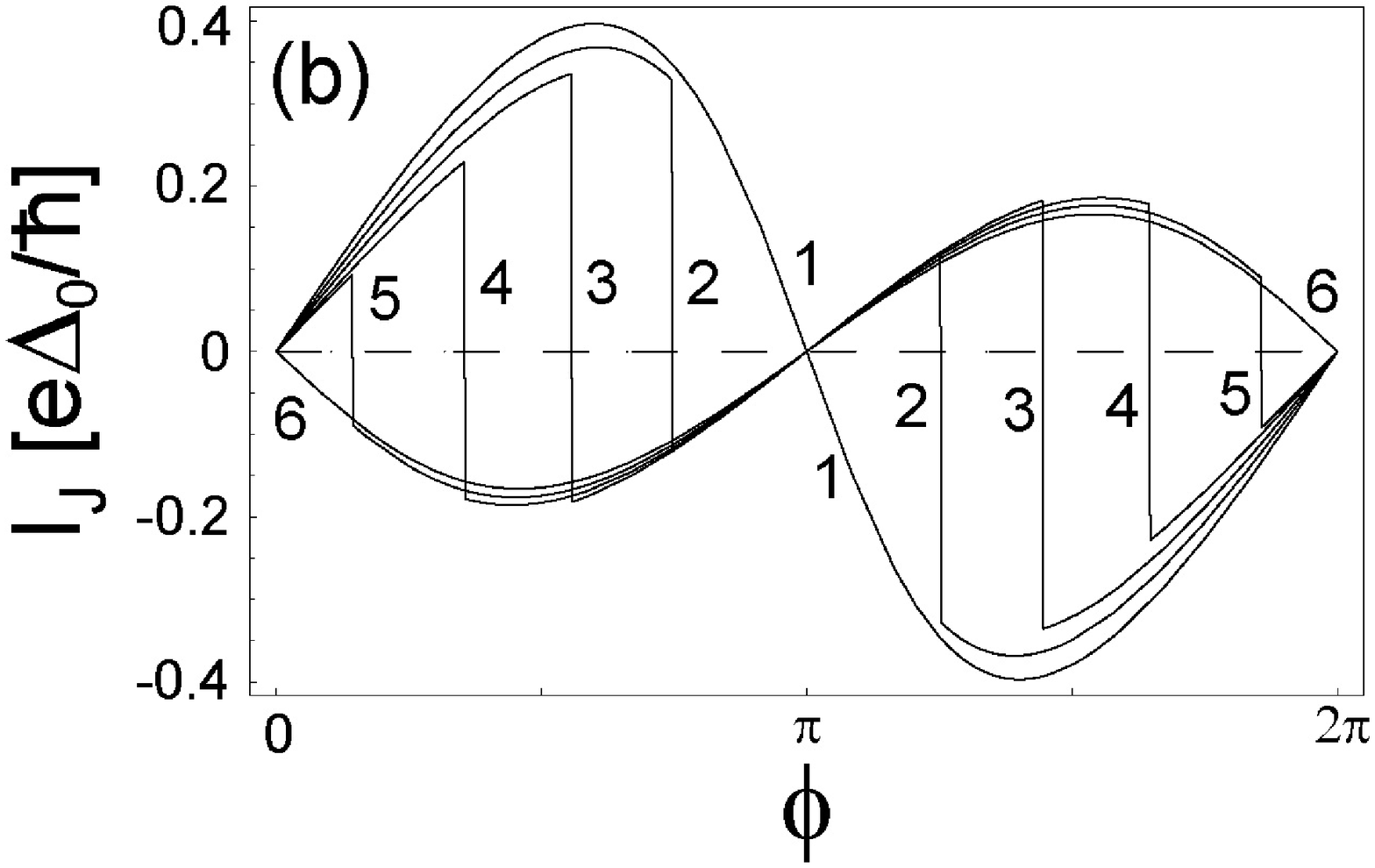}
\caption{\label{iccontours} (a) Behavior of $I_J(\phi)$ in the
$(g,Z)$-plane. White regions: $I_J$ is continuous. Light (dark) grey
regions: Maximum of $|I_J(\phi)|$ is realized at a continuous
extremum (discontinuity). Black area: Only extrema at
discontinuities exist. The white regions are bounded by $\Z^2=g^2$
and by $\Z^2=g^2-1$. (b) $I_J(\phi)$ at $T=0$ for the six parameter
pairs $(g,\Z)$ indicated by dots in (a).}
\end{figure}

While the Josephson current is a continuous function of $\phi$ for
certain combinations of $Z$ and $g$, $I_J$ also exhibits
discontinuities, in particular in the vicinity of $Z \approx g$, as
shown in Fig.~\ref{EandIj}(b). These discontinuities arise from a
zero energy crossing of an Andreev state at a certain phase,
$\phi_{LC}$, where $\p E_{\al,\be}/\p \phi \not = 0$, as shown in
Fig.~\ref{EandIj}(a). Since at $T=0$, only the negative energy
branches of the bound states are populated and thus contribute to
$I_J$, $\p E_{\al,\be}/\p\phi\not=0$ leads to a discontinuity in the
Josephson current at $\phi_{LC}$. In Fig.~\ref{iccontours}(a), we
present a contour plot that represents in which parts of the
$(g,Z)$-plane, $I_J$ exhibits a continuous or discontinuous
dependence on $\phi$. In the white regions of
Fig.~\ref{iccontours}(a), $I_J$ is continuous, while in the grey and
black regions, which are located in the vicinity of the $Z=g$ line,
it exhibits discontinuities as $\phi$ is varied.

This change between continuous and discontinuous behavior of $I_J$
in the $(g,Z)$-plane leads to an interesting form of the critical
current, $I_c$. Here, we define $I_c$ for a given $g$ and $\Z$ as
the maximum absolute value of $I_J$ for any $0\leq\phi<\pi$, i.e.,
$I_c=\max_{\phi}\left[|I_J(\phi)|\right]$. At that value of $\phi$,
for which $|I_J(\phi)|$ exhibits the maximum value, $I_J$ either
possesses either a continuous extremum or a discontinuity. In the light
(dark) grey areas of Fig.~\ref{iccontours}(a), $I_c$ is realized at
a continuous extremum (discontinuity), while the discontinuity
(continuous extremum) realizes a local maximum of $|I_J|$ only. In
the black regions of Fig.~\ref{iccontours}(a), no continuous
extremum exists. In order to investigate the origin of the
qualitatively different behavior of $I_J$ in the white, grey and
dark regions, we consider six pairs of values $(g,Z)$, denoted by
the dots in Fig.~\ref{iccontours}(a), and present the resulting
Josephson currents as a function of $\phi$ in
Fig.~\ref{iccontours}(b). We find that the different behavior in the
grey and black regions arises from a shift in the values of $\phi$
at which the discontinuity in $I_J$ occurs. As one moves from point
(1) to (6) in Fig.~\ref{iccontours} two discontinuities first emerge
at $\phi=\pi$, and then move symmetrically toward $\phi=0$ and
$\phi=2\pi$, respectively.
%
% Figure 2
%

\begin{figure}[ht]
\includegraphics[width=8.cm,angle=0]{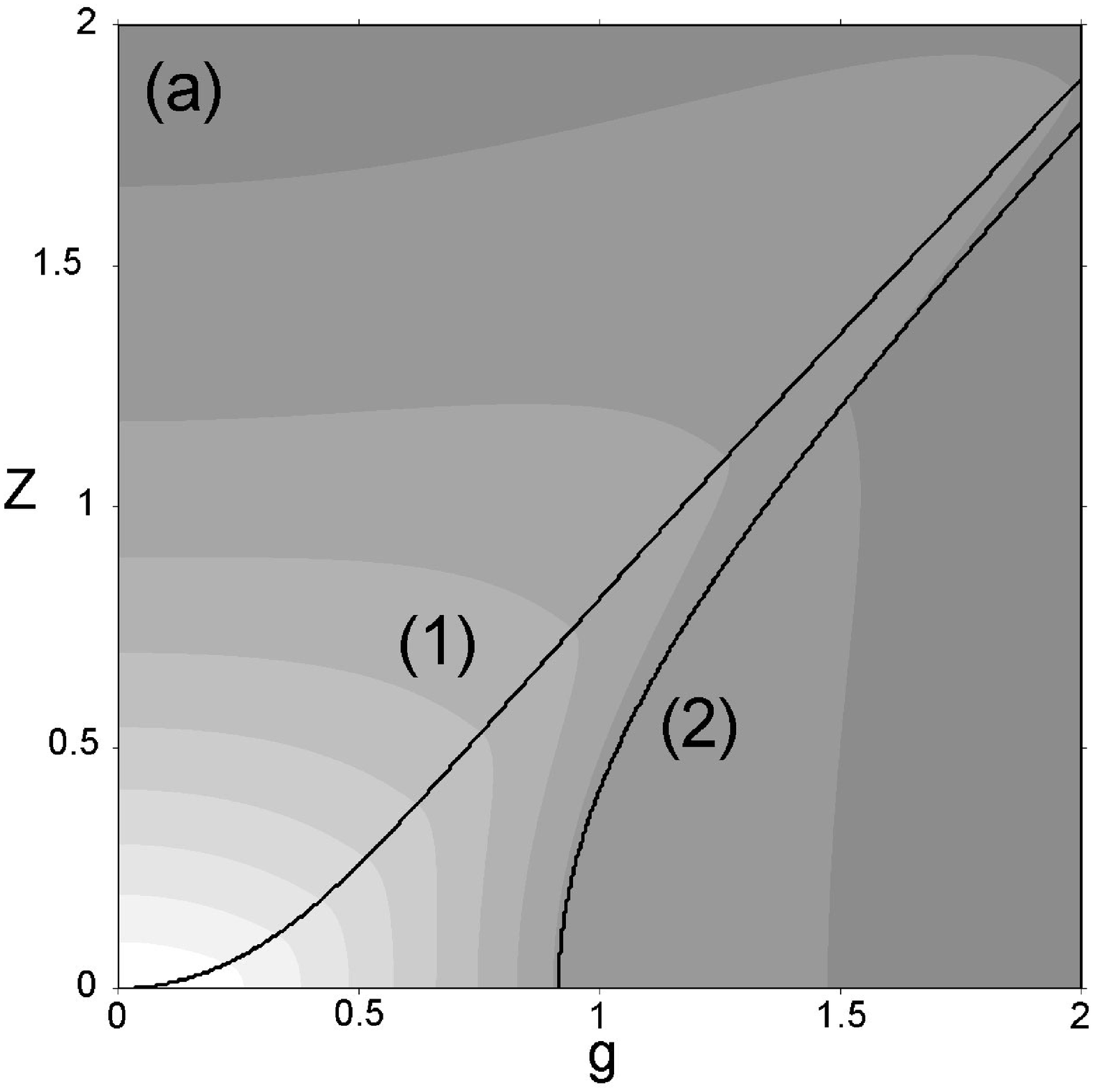}
\includegraphics[width=8.cm,angle=0]{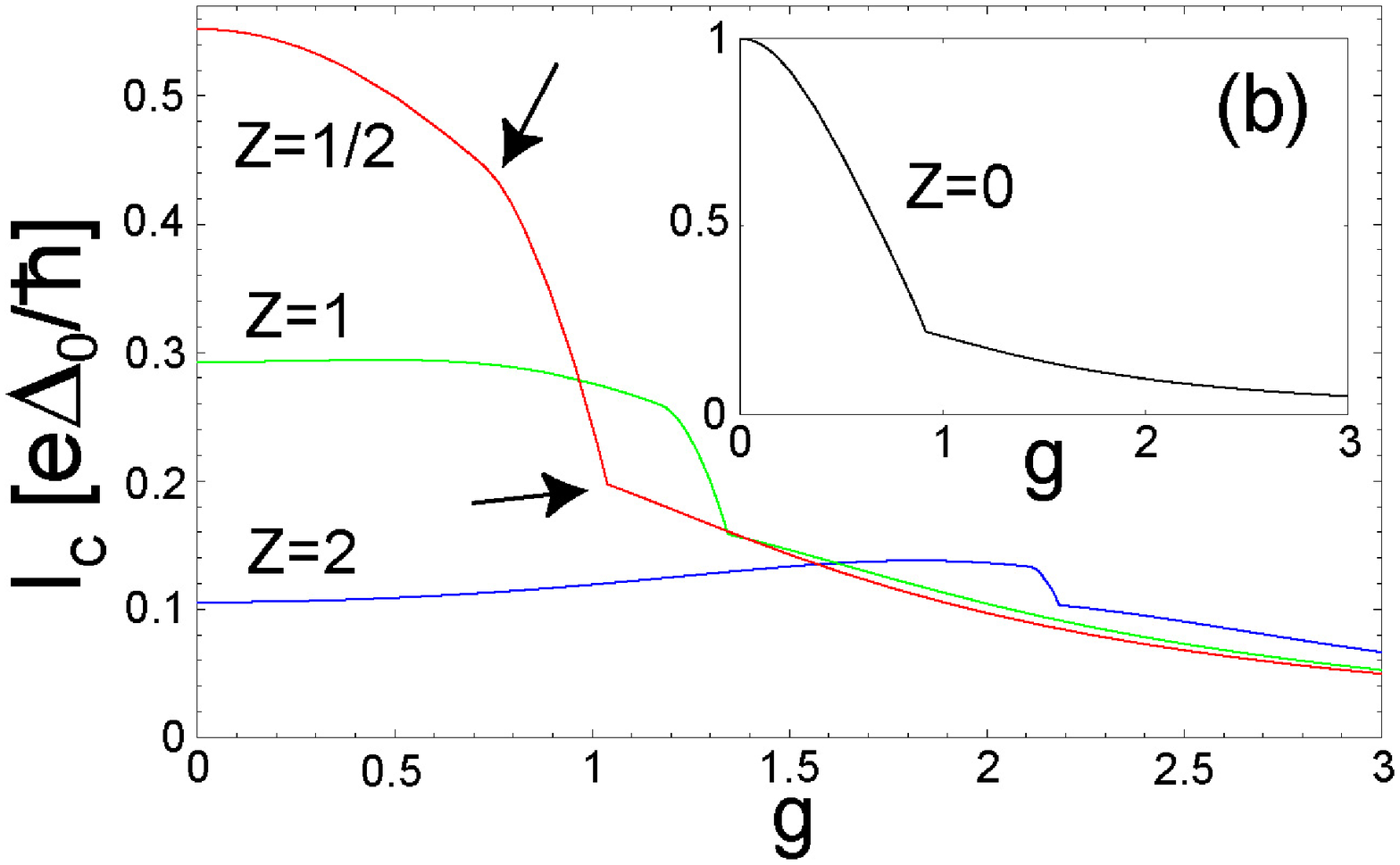}
\includegraphics[width=8.cm,angle=0]{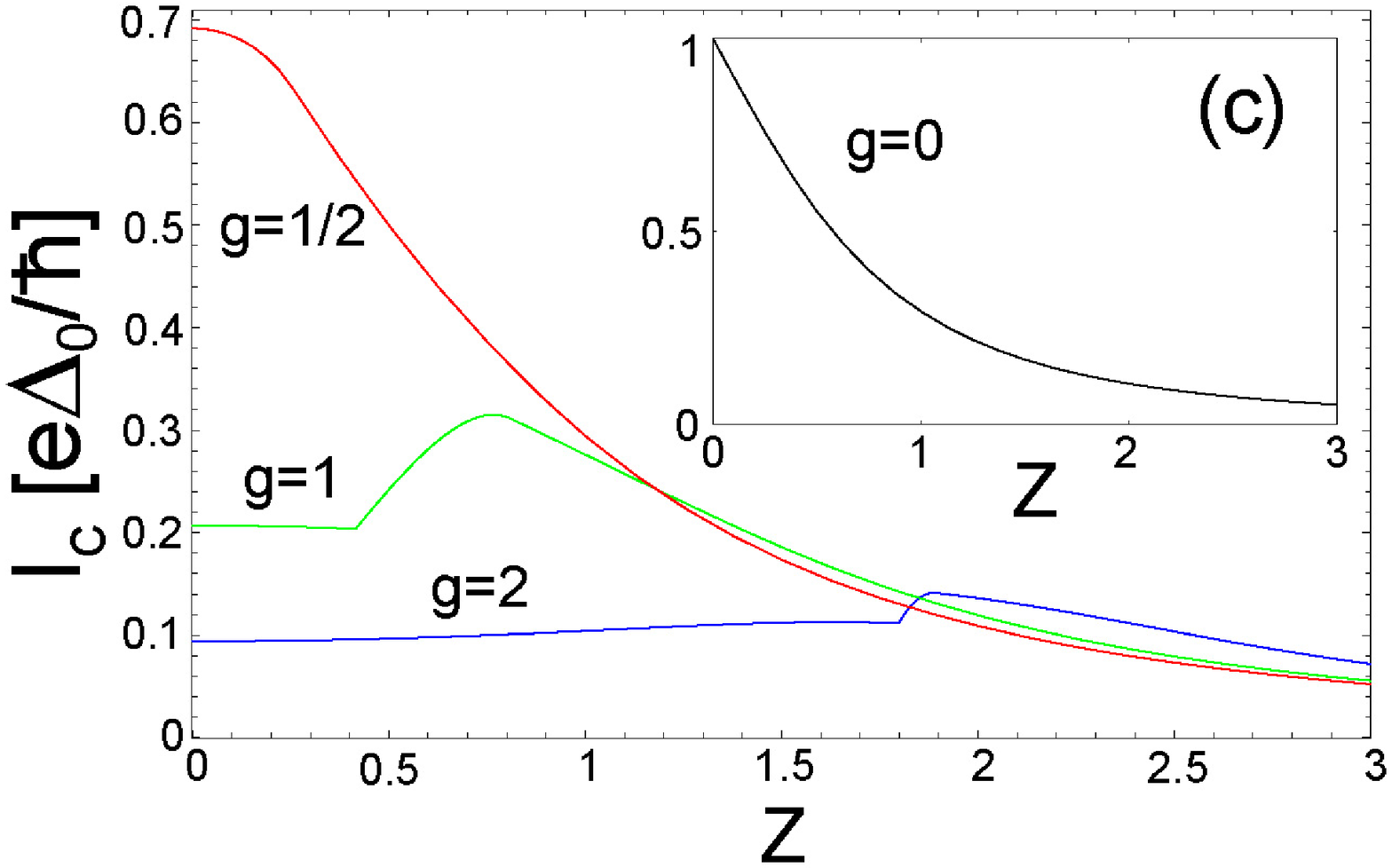}
\caption{\label{ic2dplots} (Color online). (a) Contour plot of $I_c$
as a function of $g$ and $\Z$ [white (black) areas indicating a
large (small) critical current]. The solid black lines indicate
non-analyticities in the critical current. (b) $I_c$ as a function
of $g$ for several values of $\Z$. (c) $I_c$ as a function of $\Z$
for several values of $g$.}
\end{figure}

In Fig.~\ref{ic2dplots}(a), we present a contour plot of the
critical current in the $(g,Z)$-plane, with white (dark grey) areas
indicating a large (small) critical current. An analytical
expression for the critical current in the $(g,Z)$-plane can be
obtained along the lines $g=0$ and for $Z=0$, where one finds
\begin{align}
\label{Ic}
I_c(g{=}0,\Z)=&\f{e\Dz}{\hbar}\left(1-\f{\Z}{\sqrt{1+\Z^2}}\right)\ , \\
I_c(g,\Z{=}0)=&
\begin{cases}
\ds \f{e\Dz}{\hbar} \f{\sqrt{1-g^2}}{1+g^2}
& 0\leq g\leq g_m,\\
\ds \f{e\Dz}{\hbar} \left( \f{g}{\sqrt{1+g^2}}-\f{g^2}{1+g^2}
\right)\!\!
&\quad g\geq g_m,
\end{cases}
\end{align}
where $g_m\approx0.915186$ is the largest real solution of the
equation $4g^6-4g^2+1=0$.

%
% Figure 3
%
Even though the critical current is a continuous function of $g$ and
$Z$, it possesses two lines of non-analyticity in the $(g,Z)$-plane
that asymptotically approach $Z=g$ for $Z,g \rightarrow \infty$.
These lines are represented in Fig.~\ref{ic2dplots}(a) as solid
black lines. Line (1), corresponding to the boundary between the
light grey and the black region in Fig.~\ref{iccontours}(a), represents
a discontinuity in the second derivative of $I_c$. In contrast, line
(2), which corresponds to the boundary between the light grey and
dark grey region in Fig.~\ref{iccontours}(a,) represents a
discontinuity in the first derivative of $I_c$. Line (2) also
represents a sign change in that value of $I_J$ which determines
$I_c$. In other words, to the left (right) of line (2), $I_c$ is
realized by a positive (negative) value of $I_J$. These
non-analyticities  become particularly apparent when one plots the
critical current as a function of $g$ (for constant $Z$) and $Z$
(for constant $g$), as shown in Figs.~\ref{ic2dplots}(b) and (c),
respectively [in Fig.~\ref{ic2dplots}(b), we indicated the
non-analyticities for the curve with $Z=1/2$ by arrows].

\begin{figure}
\includegraphics[width=8.cm,angle=0]{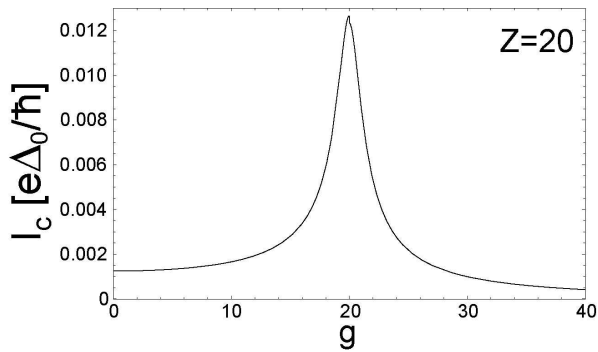}
\caption{\label{gZ} $I_c$ as a function of $g$ for $Z=20$. The plot
of $I_c$ as a function of $\Z$ for $g=20$ is virtually
indistinguishable from this plot.}
\end{figure}

The different dependence of $I_J$ on the scattering strength in
the limits $g \gg Z$ or $Z \gg g$ [see Eqs.~(\ref{IJnmag}) and
(\ref{IJmag})] on one hand and $g=Z \gg1$,
where
\begin{subequations}
\label{IJZeqg}
\begin{align}
I_J^\alpha=&\f{e \Dz}{\hbar}\f{\sin \phi}{4Z}\ ,\\
I_J^\beta=&\f{e \Dz}{\hbar}\f{\sin \phi
\left(1-\cos\phi\right)}{16Z^2} \ ,
\end{align}
\end{subequations}
% [see
%Eq.~(\ref{IJZeqg})]
on the other hand leads to the interesting possibility to increase
the critical current by increasing the magnetization, and hence the
scattering strength of the barrier. Specifically, we find $I_c=\f{e
\Dz}{2\hbar}g^{-2}$ for $g \gg Z$ and $I_c=\f{e \Dz}{2\hbar}Z^{-2}$
for $Z \gg g$, while for $Z=g \gg 1$, we have $I_c=\f{e
\Dz}{4\hbar}Z^{-1}$. As a result, we find that for fixed $Z \gg 1$
and increasing $g$, the critical current exhibits a maximum at
$Z=g$. This effect is demonstrated in Fig.~\ref{gZ} where we present
$I_c$ as a function of $g$ for $Z=20$.
%
% Figure 4
%
Hence, for a given non-magnetic scattering strength of a
paramagnetic barrier, it is possible to increase the critical
current by increasing the magnetization of the barrier and thus its
magnetic scattering strength. This increase of the magnetization
can, for example, be achieved by applying a local magnetic field via
atomic force microscopy \cite{Ham04}.

\begin{figure}[ht]
\includegraphics[width=8.cm,angle=0]{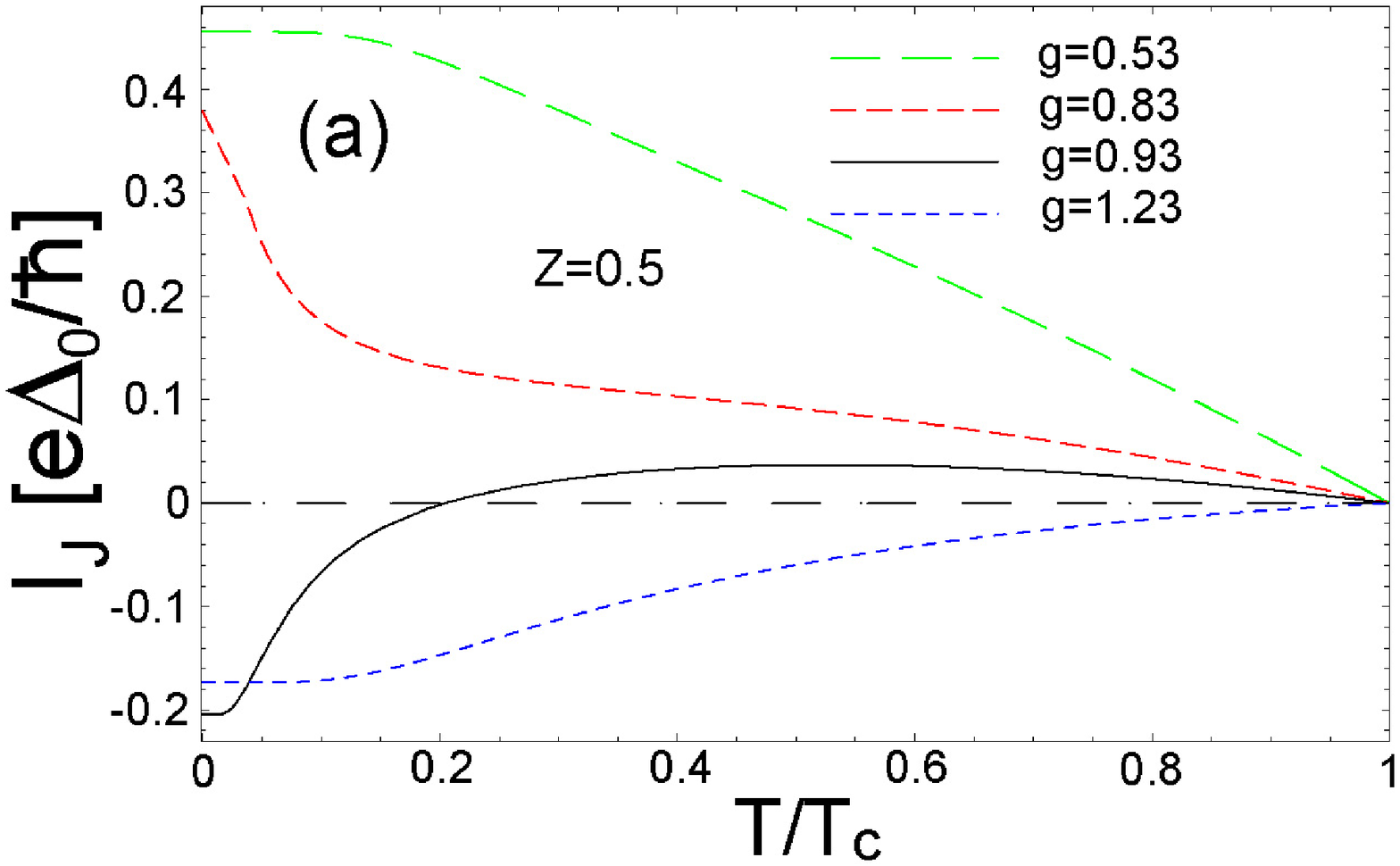}
\includegraphics[width=8.cm,angle=0]{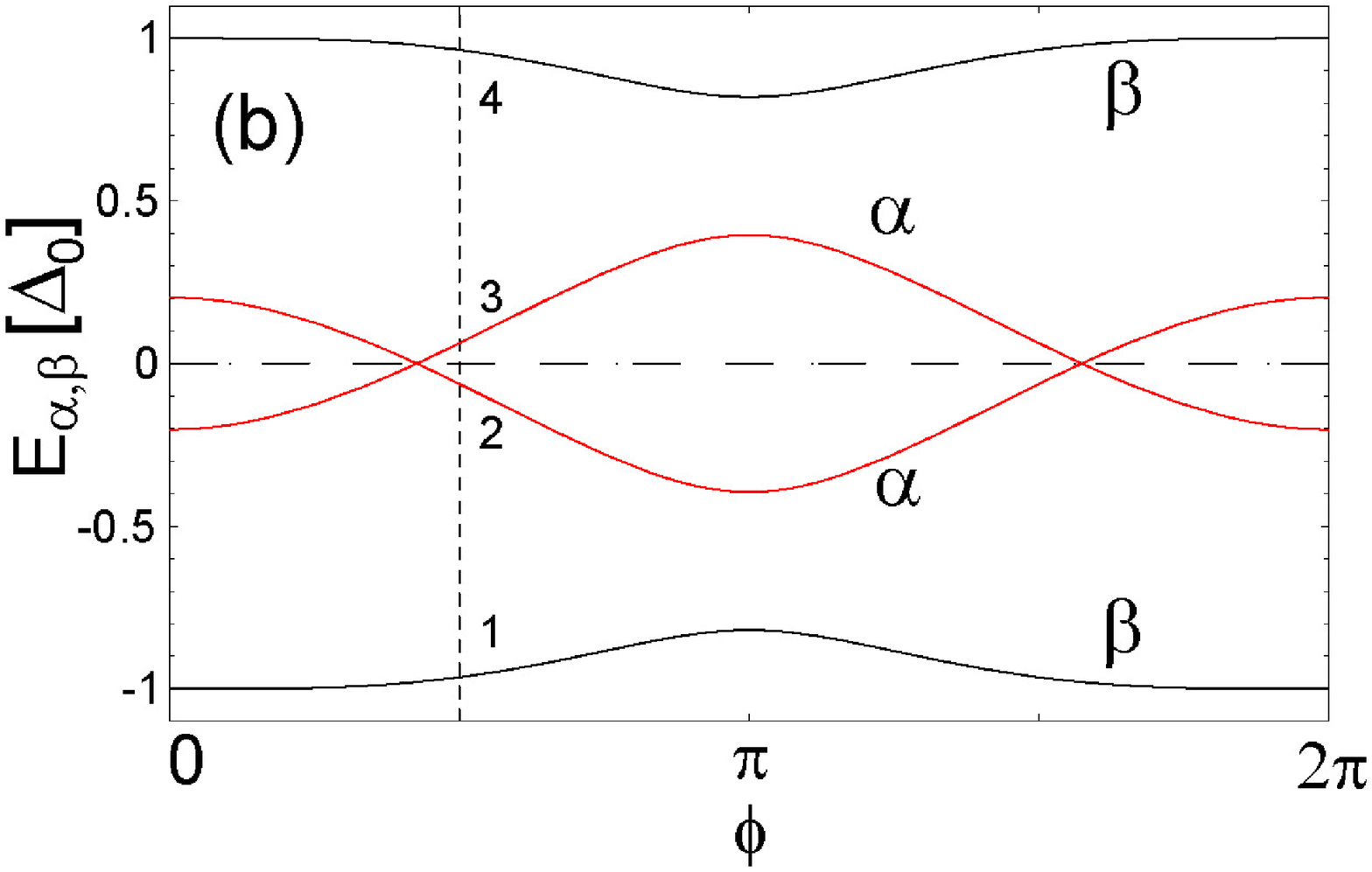}
\caption{\label{IJT} (Color online). (a) $I_J$ as a function of
$T/T_c$ for $\phi=\f{\pi}{2}$, $\Z=1/2$ and several values of $g$.
(b) $\pm E_{\al,\be}$ as a function of $\phi$ for $\Z=1/2$ and
$g=0.93$. The dotted line corresponds to the case $\phi=\f{\pi}{2}$
of (a).}
\end{figure}

We next consider the temperature dependence of the Josephson
current. Here, we find that the splitting of the Andreev states for
a non-zero magnetic scattering strength can lead to an
unconventional temperature dependence of  $I_J$ in which it changes
sign with increasing temperature {\it without} a change in the
relative phase, $\phi$, between the two superconductors. This
temperature dependence is demonstrated in Fig.~\ref{IJT}(a), where
we assume a BCS temperature dependence of the superconducting gap.
%
% Figure 5
%
In order to understand this sign change, we consider the
$\phi$-dependence of $E_{\alpha,\beta}$ which is shown in
Fig.~\ref{IJT}(b). At $T=0$, only the branches indicated by $1$ and
$2$, belonging to Andreev states $\beta$ and $\alpha$, respectively,
are occupied. Since the derivatives $\p E_{\alpha}/\p \phi$ and $\p
E_{\beta}/\p \phi$ possess opposite signs, the Josephson currents
through them, $I_J^\al<0$ and $I_J^\be>0$, flow in opposite
directions with $|I_J^\al| > |I_J^\be|$. Since with increasing
temperature, the occupation of branches $2$ and $3$ changes more
rapidly than those of branches $1$ and $4$, it follows that the
magnitude of $I_J^\beta$ decreases more quickly than that of
$I_J^\alpha$. As a result, the total current,
$I_J=I_J^\alpha+I_J^\beta$, eventually changes sign. A possible sign
change of $I_J$ with increasing temperature was previously also
discussed in Refs.~\onlinecite{Fog00,Bar02}. However, due to the
differences between our results for $E_{\alpha,\beta}$ and those in
Refs.~\onlinecite{Fog00,Bar02} [see also Ref.~\onlinecite{com1}] it
is presently unclear, whether the origin of the sign change in
Refs.~\onlinecite{Fog00,Bar02} is the same as the one discussed
here.

The qualitative nature of the temperature dependence can be altered
via a change of the couplings $g$ and $\Z$, as follows from
Fig.~\ref{IJT}(a). We find that in general, a sign change of $I_J$
with increasing temperature occurs when (a) the particle-like
components of both Andreev states possess the same spin-polarization
and (b) $\phi$ is chosen such that the energy difference between the
Andreev states is sufficiently large. These two conditions can
easily be satisfied if at least one of the Andreev states exhibits a
zero-energy crossing, and $\phi$ is chosen to be close to that
crossing in the region where $\langle S_z \rangle = 1/2$. A zero
energy crossing, however, occurs only in the grey and black regions
of the $(g,Z)$-plane shown in Fig.~\ref{iccontours}(a). Thus, in
order to observe a sign change of $I_J$ with temperature, one should
select a barrier whose scattering potentials are close to the $Z=g$
line. Note that a similar temperature dependent sign change is also
predicted to occur in Josephson junctions consisting of triplet
superconductors and a ferromagnetic barrier \cite{tft}. It is
important to stress that the temperature dependent sign change
discussed above is {\it qualitatively} different from the one
reported by Ryazanov {\it et al.} \cite{Rya01}. There, the sign
change arises from a transition of the junction from a $0$-phase
state at high temperatures to a $\pi$-phase state at low
temperatures due to a temperature dependent coherence length
\cite{reviews}. In contrast, in our case the sign change arises from
a change in the population of the Andreev states, with the relative
phase between the superconductors remaining unchanged.

\subsection{Spin Transport through Andreev States}
\label{sec:spintransport}

\begin{figure}
\includegraphics[width=7.5cm,angle=0]{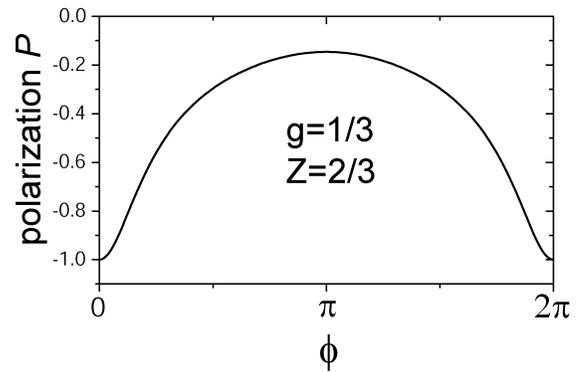}
\caption{\label{fig:polphi1} Spin polarization ${\cal P}$ of the
Josephson current through the Andreev states at $T=0$ as a function
of $\phi$ for $g=1/3$ and $Z=2/3$. }
\end{figure}

We argued in Sec.~\ref{QMmethod}, that the origin of the zero total
spin current lies in the fact that the spin current through the
Andreev states is compensated by a spin current through the
continuum states that is equal in magnitude, but opposite in sign.
The question thus naturally arises if it is possible to separately
measure these two contributions to the total spin current. While one
could envision several experimental set-ups in which this could be
achieved, for example, by using two Josephson tunneling STM tips,
one on each side of the junction, we cannot provide a definite
answer to this question at the moment. However, the ability to
measure these contributions separately would open new venues for
using the combined spin and charge degrees of freedom in such a
junction. In particular, it would be possible to make use either of
a spin polarized (non-zero) charge Josephson current by considering
the current through the Andreev states, or of a spin Josephson
current without a charge Josephson current by considering the
current through the continuum states. To exemplify these
possibilities, we consider in what follows the spin polarization of
the Josephson current through the Andreev states. We first define a
spin polarization ${\cal P}$ of the Josephson current via
\begin{eqnarray}
{\cal P}  = \frac{I^{AS}_\uparrow(0) - I^{AS}_\downarrow
(0)}{I^{AS}_\uparrow (0)+ I^{AS}_\downarrow (0)} = -e
\frac{I^{AS}_\uparrow (0)- I^{AS}_\downarrow (0)}{I_J} \ .
\label{eq:pol}
\end{eqnarray}
For ${\cal P}=-1 (+1)$, the Josephson current through the Andreev
states is completely spin polarized and thus solely carried by
\mbox{spin-$\da$} (\mbox{spin-$\ua$}) electrons. In
Fig.~\ref{fig:polphi1} we present ${\cal P}$ as a function of $\phi$
for $g=1/3$, $Z=2/3$ and $T=0$ (the corresponding charge Josephson
current is shown in the second panel of the second row in
Fig.~\ref{EandIj}b). We find that ${\cal P}$ is non-zero for all
$\phi$, and its magnitude reaches a maximum at $\phi=\pi/2$ (note
that while ${\cal P} \rightarrow-1$ at $\phi \rightarrow 0$, one has
at the same time $I_J \rightarrow 0$).
%
% Figure 1.1
%
At $T=0$, $I^{AS}_\ua$ ($I^{AS}_\da$) is carried solely by the
occupied particle-like branch of the $\beta$-state ($\alpha$-state),
as follows immediately from Eq.~(\ref{eq:IAS}). This is consistent
with the observation in Eq.~(\ref{eq:Ak}) that the particle-like
component of the $\alpha$ and $\beta$ states possess the spin
quantum number $S_z=-1/2$ (\mbox{spin-$\da$}) and $S_z=+1/2$
(\mbox{spin-$\ua$}), respectively. Since
$|I^v_{0,\alpha}|>|I^v_{0,\beta}|$ for all $\phi$, one finds that
the Josephson current through the Andreev states is partially
\mbox{spin-$\da$} polarized, as shown in Fig.~\ref{fig:polphi1}. The
degree of spin-polarization varies with $\phi$ due to the changing
relative contributions of $I^v_{0,\alpha}$ and $I^v_{0,\be}$ to
$I_J$. Finally, for $g=0$, the two Andreev states are degenerate,
and hence ${\cal P}=0$.

\subsection{First Order Quantum Phase Transition}
\label{sec:firstorder}

Another interesting effect arising from the combination of magnetic
and non-magnetic scattering strength of the barrier is the
possibility to tune the Josephson junction through a first-order
quantum phase transition in which the ground state of the entire
system (i.e., the combined ground state of both superconductors)
changes its spin polarization from $\langle S_z \rangle =0$ to
$\langle S_z \rangle =1/2$ (assuming without loss of generality that
the magnetization of the barrier points along the ${\hat z}$-axis).
This type of first order transition, which is well known from static
magnetic impurities in $s$-wave superconductors, where it was first
discussed by Sakurai \cite{Sak70}, can occur either with increasing
scattering strength, $J$, of a magnetic impurity\cite{Sak70}, or due
to quantum interference effects \cite{Morr03}; its fingerprint is a
zero-energy crossing of the impurity induced fermionic states inside
the superconducting gap. The phase transition arises from a level
crossing in the superconductor's free energy, ${\cal F}$, resulting
in a discontinuity of $\partial {\cal F}/\partial J$ at the
transition; hence the first order nature of the transition (for a
more detailed discussion see Ref.~\onlinecite{Balatsky_RMP}). The
Josephson junction considered here provides a new possibility to
tune the system through such a first-order phase transition by
varying the phase difference between the superconductors. In
Appendix \ref{sec:junctionspin}, we explicitly show that the change
in the spin polarization of the junction coincides with the zero
energy crossing of the $\alpha$ state, and is thus solely driven by
the Andreev states. In contrast, the contribution of the continuum
states to the spin polarization vanishes.

As an example of such a phase transition we consider the case
$Z=1/3$ and $g=2/3$, for which the energies of the Andreev states
and the resulting Josephson current are shown in the third panel in
the third row of Figs.~\ref{EandIj}(a) and (b), respectively [we
indicate in the panel in Fig.~\ref{EandIj}(a) the spin quantum
number of all the components of the Andreev states]. With increasing
$\phi$, the $\alpha$ state crosses zero energy at $\phi_c^\alpha
=2\arccos{\sqrt{g^2-\Z^2}} \approx 0.608 \pi$, such that for $\phi
> \phi_c^\alpha $, the \mbox{spin-$\ua$} component of the $\alpha$ state
is particle-like, while its \mbox{spin-$\da$} component is
hole-like. This transition results in a change of the ground state
spin polarization from $\langle S_z \rangle =0$ to $\langle S_z
\rangle =1/2$.  Moreover, the $\alpha$ state crosses zero-energy
again at ${\phi_c^\alpha}' =2[\pi-\arccos{\sqrt{g^2-\Z^2}}] \approx
1.392 \pi$, such that for ${\phi > \phi_c^\alpha}'$ its
\mbox{spin-$\ua$} component is hole-like, while its
\mbox{spin-$\da$} component is particle-like. As a result, the spin
ground state changes from $\langle S_z \rangle =1/2$ back to
$\langle S_z \rangle =0$. The range of $\phi$ for which $\langle S_z
\rangle =1/2$ is indicated in the panel of Fig.~\ref{EandIj}(b) by
dotted lines. In general, one finds that $\langle S_z \rangle =1/2$
for those $\phi$ which satisfy $\cos^2(\phi/2)<g^2-Z^2$.
Consequently, as $g$ is further increased (keeping $Z$ fixed), the
range of $\phi$ for which $\langle S_z \rangle =1/2$ increases. When
$g$ exceeds the upper critical value $g_c^>=\sqrt{1+\Z^2}$, one
finds $\langle S_z \rangle =1/2$ for all $\phi$. In contrast, when
$g$ is smaller than the lower critical value $g_c^<=\Z$, one has
$\langle S_z \rangle = 0$ for all $\phi$. As already mentioned
above, the total spin current through the junction is zero even in a
state with $\langle S_z \rangle =1/2$.

\section{Conclusions}

In summary, we have studied the Josephson current, $I_J$, in a 1D
Josephson junction consisting of two $s$-wave superconductors and a
thin ($\delta$-function type) ferromagnetic barrier.  To this end,
we used two complementary theoretical approaches: the BTK method,
and an approach starting from the quantum mechanical definition of
the current operator. We discussed the general dependence of the
charge Josephson current on $g$, $Z$, and the relative phase,
$\phi$, between the two superconductors. Specifically, we showed
that in certain regions of the $(g, Z)$-plane, $I_J$ varies
continuously with $\phi$, while in other regions, and particularly
around $Z=g$, $I_J$ exhibits discontinuities. We computed the
critical current, $I_c$, defined as the maximum Josephson current
for a given $g$ and $Z$, and we showed that it possesses two lines of
non-analytic behavior in the $(g, Z)$-plane. These non-analyticities
correspond to discontinuities in the first and second derivative of
$I_c$ (with respect to $g$ or $Z$). We demonstrated that $I_c$
exhibits qualitatively different dependencies on the scattering
strength in different parts of the $(g, Z)$-plane, which opens the
interesting possibilities to increase the critical current through
the junction by increasing the junction's magnetization. This effect
possesses potential applications in the fields of quantum
information technology \cite{Kane98}. Moreover, we showed that for
certain values of $g, Z$, the Josephson current changes sign (and
thus direction) with increasing temperature without a change in the
relative phase between the two superconductors, i.e., without a
transition between a $0$ and $\pi$ state of the junction. We showed
that this sign change is entirely due to a temperature-dependent
change in the occupation of the Andreev states. In agreement with
earlier results, we demonstrated that the continuum states do not
contribute to the charge current, and that therefore the charge
Josephson current is carried entirely by the Andreev states. We also
showed that while the total spin Josephson current through the
junction is zero, the Andreev states and the continuum states
separately carry a non-zero spin current of equal magnitude but
opposite sign. The possibility of measuring these contributions
separately would open new venues for employing the combined spin and
charge degrees of freedom in such a junction for potential
applications in spin electronics and quantum information technology.
Finally, we demonstrate that by changing the phase $\phi$ between
the superconductors, it is possible to tune the junction through a
first-order quantum phase transition in which the spin polarization
of the superconductors' ground state changes between $\langle S_z
\rangle =0$ and $\langle S_z \rangle =1/2$.

Experimentally, the effects discussed here could be studied in
junctions with magnetically doped insulating barriers based on MgO,
ZnO, or TiO$_2$. In these materials one can imagine to vary $g$
independently via the substitution of magnetic dopants such as Co, Mn,
etc.~and/or by changing their concentration, or by applying a small
magnetic field, for example, via atomic force microscopy
\cite{Ham04}. Moreover, $Z$ can be altered by the choice of material
and the junction width. It is possible to control the barrier width
of complex oxides using layer-by-layer growth techniques monitored
by reflection high energy electron diffraction (RHEED) on the
unit-cell level, which is much smaller than the coherence length of
a typical $s$-wave superconductor. Hence, we expect that the results
derived above for a $\delta$-functional barrier should be observable
in experimental systems with a non-zero barrier width, $d$, as long
as $d$ is much smaller than the superconducting coherence length.
Note that the novel Josephson behavior described in this paper will
occur in addition to the effects that are expected from the
proximity induced sign change in the superconducting order parameter
as a function of the ferromagnetic barrier thickness \cite{Rya01}.

Finally, scattering off the barrier leads to a suppression of the
superconducting order parameter near the barrier, which was not
accounted for in the approach presented above. However, in $s$-wave
superconductors, the length scale over which the order parameter
recovers its bulk value near scattering centers (such as a junction)
is set by $1/k_F$ \cite{suppression}. This length scale is in
general much shorter than both the superconducting coherence length
$\xi_c $ and the decay length of the Andreev bound states,
$\kappa^{-1}$, with $\xi_c\leq \kappa^{-1}$~~\cite{Kwon04,Zag98},
and we thus expect that the inclusion of a spatially varying order
parameter does not alter the qualitative nature of our results
presented above.

\section{Acknowledgements}

D.K.M. acknowledges financial support by the Alexander von Humboldt
Foundation, the National Science Foundation under Grant No.
DMR-0513415 and the U.S. Department of Energy under Award No.
DE-FG02-05ER46225. B.K. acknowledges financial support by DLR
(German Aerospace Center). We would like to acknowledge helpful
discussions with P.~Fulde, M.~Sigrist, M.~Kuli\'c, F.~Nogueira,
I.~Eremin, Y.S.~Barash, M.~Fogelstr\"{o}m, P.~Brydon, and D.~Manske.
Moreover, we are grateful to J.~Michelsen, V.S.~Shumeiko, and G.~Wendin
for useful communications.

\appendix

\section{Contribution of the continuum states to $I_J$ and $I_S$}
\label{sec:CSdiag}

For the scattering states' wave-function of Eq.~(\ref{Psi})
we make the ansatz
\beq
\label{CSansatz}
\Psi_{n,j}(z) =
\sum_{\de,\ep=\pm} \bpm u_{n,j,s,\de,\ep}\\v_{n,j,s,\de,\ep}\epm
e^{i(\de k_F+\ep q) z}\ ,
\eeq
with $s=L,R$ referring to the left \mbox{($z<0$)} and right
\mbox{($z>0$)} hand side of the junction, respectively, with
$j=\al,\be$, with $q>0$, and where $k_F$ is the Fermi momentum.
The corresponding solutions of Eq.~(\ref{BdG}) are subject to the
boundary conditions (\ref{BC}).
For a given $E_q>|\De|$, there are eight continuum states with a
positive- and a negative-energy branch for each. For energies small
compared to the Fermi energy, $q$ is given by (\ref{E2supergapS}).

Define
\begin{align}
\cosh\ga=\f{E_q}{\Dz}.
\end{align}
To be specific, consider the negative-energy branches with energy $-E_q$.
The BdG equations imply
\begin{align}
\label{ydef}
y_{j,s,\de,\ep}
&\equiv
e^{\de\ep\ga/2}u_{j,s,\de,\ep}
=e^{i\phi_s-\de\ep\ga/2}v_{j,s,\de,\ep}\ ,
\end{align}
with $j=\al,\be$ and $s=L,R$ and where we have omitted the index $n$
labeling the different states of energy $-E_q$.

With $\phi_L=0$ and $\phi_R=-\phi$ and defining
\begin{align}
e^{i\tau_\pm}&\equiv\f{1+i(\Z\pm g)}{\sqrt{1+(\Z\pm g)^2}},
\end{align}
so that
\begin{align}
-\f{\pi}{2}<\tau_\pm<\f{\pi}{2}
\end{align}
and
\bse
\begin{align}
Z&=\f{1}{2}\left(\tan\tau_++\tan\tau_-\right),\\
g&=\f{1}{2}\left(\tan\tau_+-\tan\tau_-\right),
\end{align}
\ese
we may organize the boundary conditions (\ref{BC}) in matrix
form by writing
\begin{align}
\label{zdef} z_j\equiv M_jy_{j,L}=M_j^\ast y_{j,R},
\end{align}
with
\begin{align}
y_{j,s}^T=(y_{j,s,+,+},y_{j,s,+,-},y_{j,s,-,+},y_{j,s,-,-})
\end{align}
and
\begin{widetext}

\begin{align}
M_{\al,\be} &= \bpm
e^{-\ga/2} & e^{+\ga/2} & e^{+\ga/2} & e^{-\ga/2} \\
e^{-i\phi/2+\ga/2} & e^{-i\phi/2-\ga/2} & e^{-i\phi/2-\ga/2}
& e^{-i\phi/2+\ga/2} \\
e^{-i\tau_\mp-\ga/2} & e^{-i\tau_\mp+\ga/2} &
-e^{i\tau_\mp+\ga/2} & -e^{i\tau_\mp-\ga/2} \\
e^{-i\tau_\pm-i\phi/2+\ga/2} & e^{-i\tau_\pm-i\phi/2-\ga/2} &
-e^{i\tau_\pm-i\phi/2-\ga/2} & -e^{i\tau_\pm-i\phi/2+\ga/2} \epm,
\end{align}
\end{widetext}
where the upper (lower) sign corresponds to $\al$ ($\be$). Thus the
negative-energy scattering states for a given energy $-E_q$ may be
represented by $y_i$. Reinstating labels $m$ and $n$ for the
different states for a given energy, we demand that for
continuum-normalized orthogonal states $y_{m,j}$ and $y_{n,j}$ holds
\begin{align}
\sum_{s=L,R}\sum_{\de,\ep=\pm}y_{m,j,s,\de,\ep}^\ast
y_{n,j,s,\de,\ep} &= z_{m,j}^\dagger Q_jz_{n,j}
\nn\\
&= \de_{mn}C_r(E_q),
\end{align}
with
\begin{align}
Q_j\equiv(M_jM_j^\dagger)^{-1}+(M_jM_j^\dagger)^{-1*}
\end{align}
and where $C_r(E_q)$ is real and may be chosen to depend on the
energy $E_q$. That is, finding an orthonormal basis of scattering
states for a given energy boils down to diagonalizing the $Q_j$ and
using (\ref{ydef}) and (\ref{zdef}) to obtain all corresponding
coefficients $u_{n,j,s,\de,\ep}$ and $v_{n,j,s,\de,\ep}$.
Numerically, we find that
\bse \label{uorth}
\begin{align}
\sum_{n(E_q)}(|u_{n,j,s,+,\pm}|^2-|u_{n,j,s,-,\mp}|^2)&=0,\\
u_{n,j,s,-,+}^\ast u_{n,j,s,+,+}-u_{n,j,s,-,-}^\ast u_{n,j,s,+,-}
&=0,
\end{align}
\ese
and the same for the $v_{n,j,s,\de,\ep}$.
Moreover, we numerically obtain
\begin{align}
\sum_{n(E_q)} \Big[u_{n,j,s,+,-}^\ast u_{n,j,s,+,+} &
-u_{n,j,s,-,-}^\ast u_{n,j,s,-,+}\Big]
\nn\\
&= d_jA_r(E_q)+ic_sA_i(E_q),
\end{align}
and the same for the $v_{n,j,s,\de,\ep}$, with $d_{\al,\be}=\pm1$,
$c_{L,R}=\pm1$, and where $A_r(E_q)$ and $A_i(E_q)$ are
dimensionless and generally nonzero real coefficient functions.

It is straightforward then to show that
\begin{align}
\label{Jfromscattstates} \sum_{n(E_q)}\text{Im} &
\Big(u_{n,\al}^*\p_zu_{n,\al}\pm v_{n,\al}^*\p_zv_{n,\al}
\nn\\
&~~ \pm u_{n,\be}^*\p_zu_{n,\be}+v_{n,\be}^*\p_zv_{n,\be}\Big)
\nn\\
&=
\begin{cases}
8kA_i(E_q)\sin2q|z|,
\\
0,
\end{cases}
\end{align}
leading immediately to the results (\ref{IuapmIda}).

\section{Spin ground state of the junction and first order phase transition}
\label{sec:junctionspin}

The total spin of the system receives contributions from both the
Andreev bound states and the scattering states. Define the spin
density by
\begin{align}
\hat{\rho}_S(z) &=
\left.\f{\hbar}{2}[\hat{\rho}_\ua(z,z')-\hat{\rho}_\da(z,z')]\right|_{z'=z},
\end{align}
with $\hat{\rho}_\si(z,z')$ from (\ref{rhodef}).
It is then straightforward to show that
\begin{align}
\lefteqn{\rho_S(z) \equiv \langle\hat{\rho}_S(z)\rangle}
\nn\\
&= \f{\hbar}{2}
\sum_n\bigg\{\left[\rho_{n,u,\al}(z)+\rho_{n,v,\al}(z)\right]
\tanh\f{\be E_{n,\al}}{2}
\nn\\
&\phantom{~~~~~~~~~~~~~}
-\left[\rho_{n,v,\be}(z)+\rho_{n,u,\be}(z)\right] \tanh\f{\be
E_{n,\be}}{2}\bigg\},
\end{align}
where
\begin{align}
\rho_{n,f,j}(z)&\equiv-\frac{1}{2}\fast_{n,j}(z)\ff_{n,j}(z),
\end{align}
with $f=u,v$ and $j=\al,\be$.

Note that for any given energy, there exist pairs of continuum
states that differ only in the spin quantum numbers of their
hole-like and particle-like branches. This result, combined with the
normalization of the continuum states then yields that their
contribution to the spin of the junction vanishes. Using next the
normalization of the bound states
\begin{align}
\int_{-\infty}^{+\infty}dz
[\uast_{0,j}(z)\uu_{0,j}(z)+\vast_{0,j}(z)\vv_{0,j}(z)]=1,
\end{align}
with $j=\al,\be$, we obtain that the spin of the system is solely
determined by the Andreev bound states and given by
\begin{align}
\label{SAS} \langle S_z\rangle &=
\int_{-\infty}^{+\infty}dz\rho_S^\text{AS}(z)
\nn\\
&= -\f{\hbar}{4} \left(\tanh\f{\be E_\al}{2}-\tanh\f{\be
E_\be}{2}\right).
\end{align}
Since our conventions are such that always $E_\be>0$, but the
sign of $E_\al$ can vary, we obtain for $T=0$
\begin{align}
\langle S_z\rangle=
\begin{cases}
0 & E_\al>0,\\
\hbar/2 & E_\al<0,
\end{cases}
\end{align}
signaling a quantum phase transition caused by the zero-energy
crossing of one of the Andreev bound states.


\begin{thebibliography}{99}
\bibitem{com1}
In Ref.~\onlinecite{Fog00} a splitting of the Andreev
states is found to occur for $\phi \not = 0$ only. The reason for
this discrepancy with the results discussed above and those in
Refs.~\onlinecite{Zik99,Bar02} lies in the particular form of the
$S$-matrix chosen in Ref.~\onlinecite{Fog00}.
\bibitem{zdef}
Note that our $\Z$ is, by definition, half the $\Z$ of
Ref.~\onlinecite{Kwon04}.
\end{thebibliography}
\end{document}